\documentclass[twocolumn]{aastex631}
\usepackage{amsmath,amssymb}

\begin{document}

\title{Radio Follow Up of a Sub-threshold GRB in the Sky Localization Area of GW241125}




\author[0009-0001-2009-4708]{Natalie Gottschlich}
\affiliation{Johns Hopkins University,
Baltimore, MD 21218, USA}

\author[0000-0001-8104-3536]{Alessandra Corsi}
\affiliation{Johns Hopkins University,
Baltimore, MD 21218, USA}

\author{S.~Bradley Cenko}
\affiliation{Astrophysics Science Division, NASA Goddard Space Flight Center, Greenbelt, MD 20771, USA}
\affiliation{Joint Space-Science Institute, University of Maryland, College Park, MD 20742, USA}
\affiliation{Department of Physics, George Washington University, 725 21st St NW, Washington, DC, 20052, USA}

\author[0000-0002-2787-1012]{Divyajyoti}
\affiliation{Gravity Exploration Institute, School of Physics and Astronomy, Cardiff University, Cardiff, CF24 3AA, United Kingdom}

\author[0000-0001-5229-1995]{James DeLaunay}
\affiliation{Department of Astronomy and Astrophysics, The Pennsylvania State University, 525 Davey Lab, University Park, PA 16802, USA}
\affiliation{Institute for Gravitation and the Cosmos, The Pennsylvania State University, University Park, PA 16802, USA}

\author[0000-0002-3714-672X]{Derek B. ~Fox}
\affiliation{Department of Astronomy and Astrophysics, The Pennsylvania State University, 525 Davey Lab, University Park, PA 16802, USA}

\author[0009-0007-1842-7028]{Tanner O'Dwyer}
\affiliation{Johns Hopkins University,
Baltimore, MD 21218, USA}

\author[0000-0003-0020-687X]{Samuele ~Ronchini}
\affiliation{Gran Sasso Science Institute (GSSI), I-67100 L'Aquila, Italy
INFN, Laboratori Nazionali del Gran Sasso, I-67100 Assergi, Italy}



\begin{abstract}

Since the \textit{Fermi} satellite’s identification of a candidate $\gamma$-ray burst (GRB) temporally coincident with GW150914, several tentative, and often debated, associations between electromagnetic (EM) transients and gravitational-wave (GW) signals from binary black hole (BBH) mergers have been reported. One such event, S241125n (later confirmed as GW241125\_010116), was identified during the fourth observing run (O4) of Advanced LIGO and found to be spatially (within the large GW localization uncertainty) and temporally coincident with a subthreshold GRB detected by the \textit{Swift} Burst Alert Telescope Gamma-ray Urgent Archiver for Novel Opportunities (BAT-GUANO). Here, we present results from a radio follow-up campaign targeting the BAT-GUANO localization region, carried out with the Karl G. Jansky Very Large Array (VLA). We also re-analyze \textit{Swift}/XRT observations of the field, and combine these results with optical upper limits. Our analysis constrains the isotropic kinetic energy of a putative relativistic jet launched in the BBH merger to $\lesssim  3\times10^{50}$\,erg for $n_\text{ISM} = 1.0$ cm$^{-3}$. We also discuss both the challenges and the diagnostic power of radio follow up in assessing candidate BBH–GRB associations, and present projections for analogous radio studies in the LIGO-Virgo-KAGRA observing run 5 (O5), and in the era of next-generation ground-based instrumentation. The enhanced sensitivity and localization capabilities of detector networks such as Cosmic Explorer and the Einstein Telescope, paired with the enhanced sensitivity of next-generation radio interferometers such as the next-generation VLA and the Square Kilometre Array, will significantly strengthen coordinated multi-messenger follow-up of BBHs. These next-generation facilities are likely to provide an answer to whether BBHs host relativistic ejecta powered by mini-disk accretion. 

\end{abstract}

\keywords{gravitational waves; radiation mechanisms: non-thermal; radio continuum: general; gamma-ray burst: individual}


\section{Introduction} \label{sec:intro}

The 2015 Nobel Prize-winning direct detection of gravitational waves (GWs) from the merger of a stellar-mass binary black hole (BBH)---GW150914---by the Advanced Laser Interferometer Gravitational Wave Observatory \citep[LIGO;][]{LIGO150914,150914Astrophys} pushed multi-messenger astronomy to a new frontier \citep{BBHLigo1,BBHLigo2,BBHLigo3}. The importance of this growing field was further demonstrated by the discovery of GW170817 \citep{170817GW}, the first binary neutron star merger discovered with GWs to have a confirmed electromagnetic (EM) counterpart \citep[and references therein]{170817GW}. 

\begin{figure}
    \centering
    \includegraphics[width=1\linewidth]{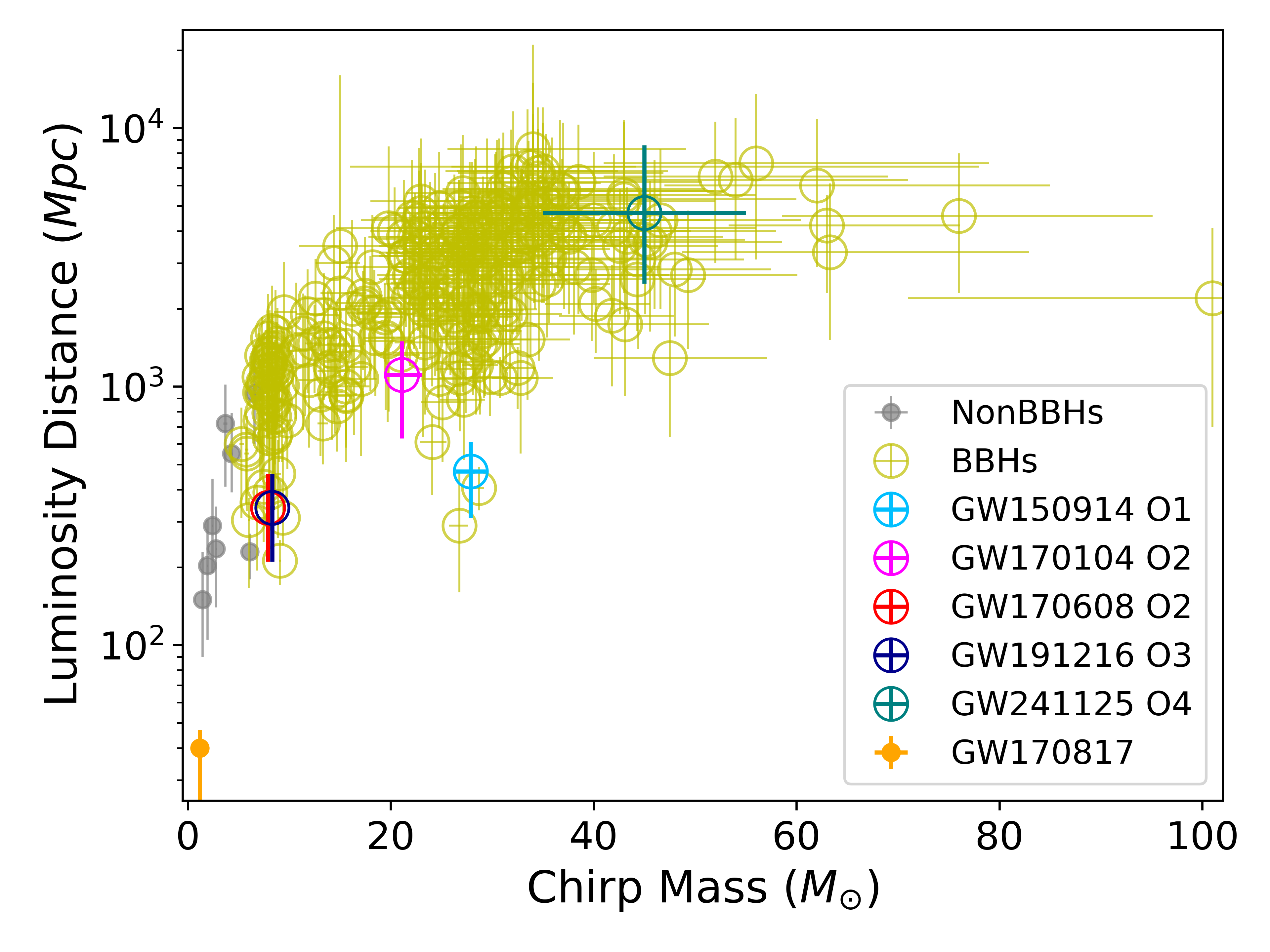}
    \caption{BBH candidates (defined as events with both component masses $\gtrsim 3 $M$_\odot$; olive) and likely BH-NS systems (secondary mass less that $3$\,M$_{\odot}$; gray) from GWTC-5 \citep{2026arXiv260527225T}. For comparison we also show the binary neutron star merger GW170817 (orange), and mark BBHs with candidate GRB associations which have emerged in every LVK observing run (O1–O4; sun crosses), including S241125n/GW241125 analyzed here.}
    \label{fig:gwtc}
\end{figure}

As the most recent observing runs of the LIGO-Virgo-KAGRA (LVK) detectors have demonstrated, the most prolific LVK sources remain stellar-mass BBHs \citep[][and references therein]{2026arXiv260527225T}. Traditional models for BBH mergers do not predict EM counterparts, expecting the surrounding environment to be free of ejecta mass and lacking the tidally disrupted material that, in the case of a neutron star-BH or neutron star-neutron star system, could provide the accretion energy necessary to power an EM-emitting relativistic outflow. However, there have been several GW events with candidate EM associations  \citep[e.g.,][and Figure \ref{fig:gwtc}]{GW170608_VLA,190521_EM}, including the case of GW150914 \citep{LIGO150914,150914Astrophys}, which was temporally coincident with a subthreshold gamma-ray burst (GRB) seen by \textit{Fermi}  \citep{FermiGW}. S241125n/GW241125\_010116  \citep{GCN38305,S241125nSummary}, the event examined here, is a candidate stellar mass BBH merger observed in the fourth observing run (O4) of the LVK detectors, found to be temporally and spatially coincident with a subthreshold GRB detected by {\it Swift}’s Burst Alert Telescope Gamma-ray Urgent Archiver for Novel Opportunities \citep[BAT-GUANO;][]{GCN38308}. 

From a theoretical standpoint, these associations have prompted debate over whether a stellar-mass BH merger can give rise to a GRB, a question that remains open as of the end of the O4 run. Theoretical models that predict EM counterparts to stellar-mass BBH mergers can be broadly divided into two classes. In the first class, the merger occurs in a dense environment like an AGN disk \citep[e.g.,][]{AGNCounterpartModel,S241125nSummary,BBH_AGN_rates}; however, these dense environments are likely to be correlated with weak radio emission \citep{AGNgasmodel, LazzatiAGNRadio}. In the second class of models, a mini-disk surrounds a stellar-mass BBH and powers relativistic jets at merger \citep[e.g.,][]{BBH_GRBs,EM_BBH_Limits,LIGO_BBH_EM}. Confirming or excluding these models would have important implications for our understanding of both stellar BH formation and of the effects of a merger's environment. 

Here, we present observations of the  sub-threshold GRB potentially associated with S241125n/GW241125\_010116  conducted with the Karl G. Jansky Very Large Array (VLA). Our observations aim to search for a potential fast-ejecta afterglow from the BBH merger. In the context of the mini-disk model, the low luminosity of the associated GRB, if real, could mean a powerful jet launched off-axis or a weaker jet on-axis. Since the radio emission is largely independent of geometric effects, radio observations can be used to constrain the kinetic energy of the ejecta and, indirectly and with large uncertainties, the mass of the accretion disk powering the jet \citep[e.g.,][]{minidisk_mass}. 

Our paper is organized as follows. In Section \ref{sec:detection}, we summarize the GW alert that unveiled the low-luminosity GRB associated with S241125n and the community's prompt EM follow-up efforts. In  Section \ref{sec:obs}, we describe EM observations that accompanied S211125n/GW241125\_010116, including our radio follow-up observations conducted with VLA, and  \textit{Swift}/XRT observations  that we re-analyze here. In Section \ref{sec:discussion} we interpret our data in the context of the mini-disk model for BBHs. In Section \ref{sec:nextgen} we present projections for analogous radio studies in the upcoming LVK intermediate run 1 (IR1) and observing run 5 (O5), and in the era of next-generation ground-based instrumentation. Finally, in Section \ref{sec:summary} we summarize and conclude.

\section{GW Trigger and Prompt EM Observations} \label{sec:detection}
On 2024 November 25 at 01:01:16.780 UTC the LVK collaboration identified the compact binary merger candidate S241125n during real-time processing of the LIGO-Virgo data \citep{GCN38305}. The candidate was found by several of the online analysis pipelines, with an estimated false alarm rate of $\approx 1/33$\, years. The initial GW localization and luminosity distance estimate placed this event within a 90\% credible region of $2367$\,deg$^2$ in the sky at $\approx 6.4\pm1.7$\,Gpc. This was later updated to a 90\% credible region of 2196 deg$^2$ at $\approx 4.2\pm1.6$\,Gpc. The fifth GW Transient Catalog (GWTC-5) re-named S241125n as GW241125\_010116 (hereafter, GW241125), and confirmed that this event has a $>98\%$ probability of being astrophysical in origin. The estimated primary ($m_1$) and secondary ($m_2$) masses for the BHs in the system are $60^{+14}_{-13}\,$M$_{\odot}$ and $47^{+14}_{-17}\,$M$_{\odot}$, respectively \citep{2026arXiv260527225T}. The GWTC-5 also revised the luminosity distance of this event to $d_L =4.7^{+3.9}_{-2.2} $\,Gpc, and the 90\% localization area to $\approx 2400$\,deg$^2$ \citep{2026arXiv260527225T}.

\begin{figure}
    \centering
    \includegraphics[width=1\linewidth]{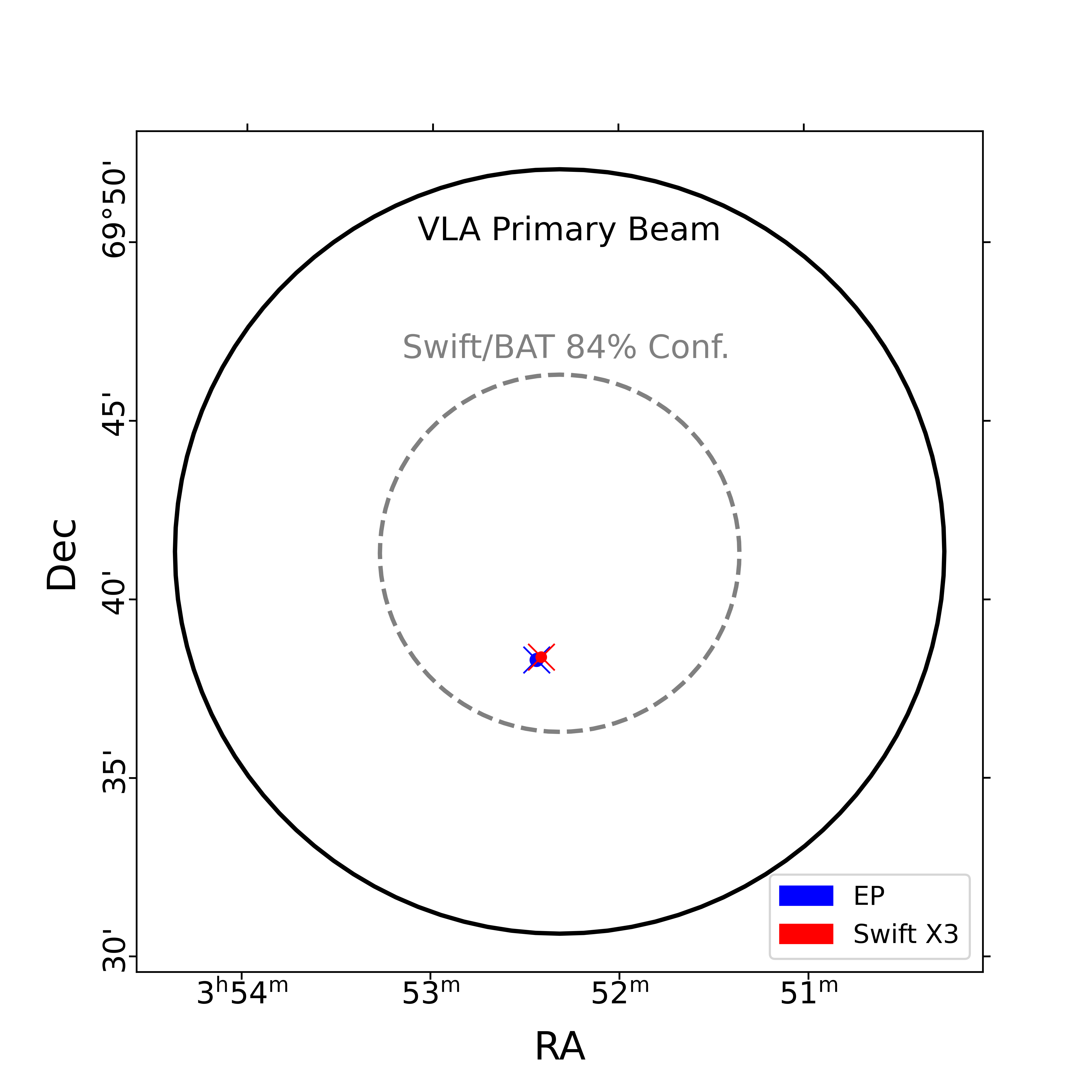}
    \caption{Localization area of the sub-threshold\textit{Swift}/BAT-GUANO transient identified during the follow-up of GW241125 (84\% of the localization area within 5\arcmin; dashed line) and 20\% of the VLA primary beam at 3\,GHz ($\approx 11\arcmin$ radius; solid line). We also mark the position of the \textit{Swift}/XRT source X3 and the EP source (see text for discussion). }
    \label{fig:regions}
\end{figure}

In response to the GW241125  alert, several observatories searched their data for potential transients located within the large GW sky localization area. The \textit{Swift} Gamma-ray Urgent Archiver for Novel Opportunities Burst Alert Telescope \citep[GUANO/BAT; ][]{GUANO_Main} found a candidate counterpart through the NITRATES analysis \citep{NITRATES} at 11.26\,s post-GW event. The initial localization was R.A., Dec. = 58.079\,deg, +69.689\,deg with an estimated 50\% containment uncertainty of $5\arcmin$ and an estimated joint GW-GRB false alarm rate of $\approx 1/12$\ years \citep{GCN38308}. Later, a revised joint GW-GRB analysis gave a combined 90\% credible region of 76 deg$^2$ with an untargeted joint false alarm rate of $\approx1/6$ years \citep{GCN38315,GCN38356}. The most recent joint localization has $\approx 84$\% of the probability within 5$'$ of the most likely position \citep[Figure\,\ref{fig:regions};][]{GCN38351}. The preliminary spectral analysis showed this $\gamma$-ray candidate's best-fit flux at $1.1\times10^{-7}$\,ergs\,cm$^{-2}$\,s$^{-1}$ \citep{GCN38351}. We derive a fluence of $\approx 5.6\times 10^{-8}$\,erg\,cm$^{-2}$ in the 15--350 keV range using the best-fit spectral template.
From 55--74 ks after the LIGO-Virgo trigger, \textit{Swift} XRT observed 0.2 deg$^2$ of the GUANO error region, and initially detected 5 uncatalogued X-ray sources, none bright enough to be strong candidate counterparts \citep{GCN38324}. This list was later updated\footnote{See \url{https://www.swift.ac.uk/LVC/S241125n/}} to include all sources in the region. 

On 2024-11-26 starting at 03:08:45 UTC, the Einstein Probe Follow-up X-ray Telescope (EP FXT) observed the position of the GUANO candidate counterpart to GW241125 for $\approx$11\,ks and detected an X-ray source within the 5$'$ error region at R.A., Dec.: 58.1097\,deg, 69.6392\,deg with an uncertainty of 10$''$ \citep{GCN38345}.  Other gamma-ray detectors observed the region of the GUANO transient and all reported non-detections. The deepest upper limits were reported by INTEGRAL, with an estimated 3$\sigma$ upper limit on the 75-2000 keV fluence of $<1.6\times10^{-7}$\,erg\,cm$^{-2}$ within the 5$'$ region for a burst lasting less than 1\,s with a characteristic short GRB spectrum \citep{GCN38311}. \textit{Fermi}/GBM, Konus-Wind, LST-1, and MAGIC each observed the same region, finding no EM counterpart candidates \citep{GCN38316,GCN38321,GCN38443}. 

Several optical telescopes also observed the GUANO region, these include Kinder, COLIBR\'I, the Himalayan Chandra Telescope, GOT, GRANDMA, GROWTH-India Telescope, and DDOTI \citep{GCN38314, GCN38317, GCN38322, GCN38325, GCN38328, GCN38334, GCN38396, GCN38329}. The deepest limits were achieved by the SAGUARO collaboration, reaching a 3$\sigma$ limit in the $r$-band of 25.5\,mag  \citep{GCN38333}.

\section{Follow-Up Observations and analysis} \label{sec:obs}

\subsection{VLA}

\begin{deluxetable}{cccc}[h]

\tablecolumns{4}
\tablewidth{1.0\columnwidth} 
\tablecaption{Summary of the VLA 3\,GHz observations of the field of the \textit{Swift}/GUANO candidate counterpart to GW241125.} 
\tablehead{ \colhead{\textbf{Epoch}} & \colhead{\textbf{Config.}}& \colhead{\textbf{Central RMS Noise}} & \colhead{$\theta_{HPBW}$} \\
\colhead{MJD} & \colhead{} & \colhead{($\mu $Jy)} & \colhead{(\arcsec)}}
\startdata
    60644.0 & A & 4.2 & 0.65 \\
    60657.9 & A & 4.2 & 0.65 \\
    60686.0 & A & 4.3 & 0.65 \\
    60850.4 & C & 5.6 & 7.0  \\
    60932.2 & B & 4.6 & 2.1  \\
\enddata  
\tablecomments{Columns in this Table are: MJD time of the observation, VLA configuration, noise RMS at the center of the image, and the nominal full-width at half-power of the VLA synthesized beam.}
\label{tab:epochs}
\end{deluxetable}

We observed the field of the \textit{Swift}/GUANO candidate counterpart to GW241125 with the VLA (Program code 23B-172; PI: Corsi) at a nominal central frequency of 3\,GHz, and with a nominal bandwidth of 2\,GHz. Our observations were carried out over five epochs (E1--E5)  between 30 November 2024  and 14 September 2025 UTC, with the VLA in its A, C, and B configurations (see Table \ref{tab:epochs}). Each epoch lasted 2.5\,hr in total, including slew for overhead and calibration.
We imaged an area centered around the position R.A.: 03h52m18.96s, Dec.: 69d41m23.42s, with a nominal 15\,$\arcmin$ FWHM primary beam, fully enclosing the 84\% confidence error area of the \textit{Swift}/GUANO 5\,$\arcmin$ localization (Figure \ref{fig:regions}).

The VLA data were calibrated using the VLA automated calibration and imaging pipeline in \texttt{CASA} \citep{2020ASPC..527..571K}. After calibration, we inspected the data for radio frequency interference (RFI) and applied any necessary flagging. Images of the field were formed using the VLA automated pipeline which includes primary beam corrections to account for the shape of the primary beam up to a region extending to 20\% of the power radius, translating to images with an angular diameter of 11\,$\arcmin$ for our observations. The non-primary beam corrected pipeline image of E3 was compared to the same image formed using the \texttt{CLEAN} algorithm in interactive mode to quantify the behavior of the imaging pipeline. Results of this comparison are included in the Appendix. 

\begin{figure}
    \centering
    \includegraphics[width=1\linewidth]{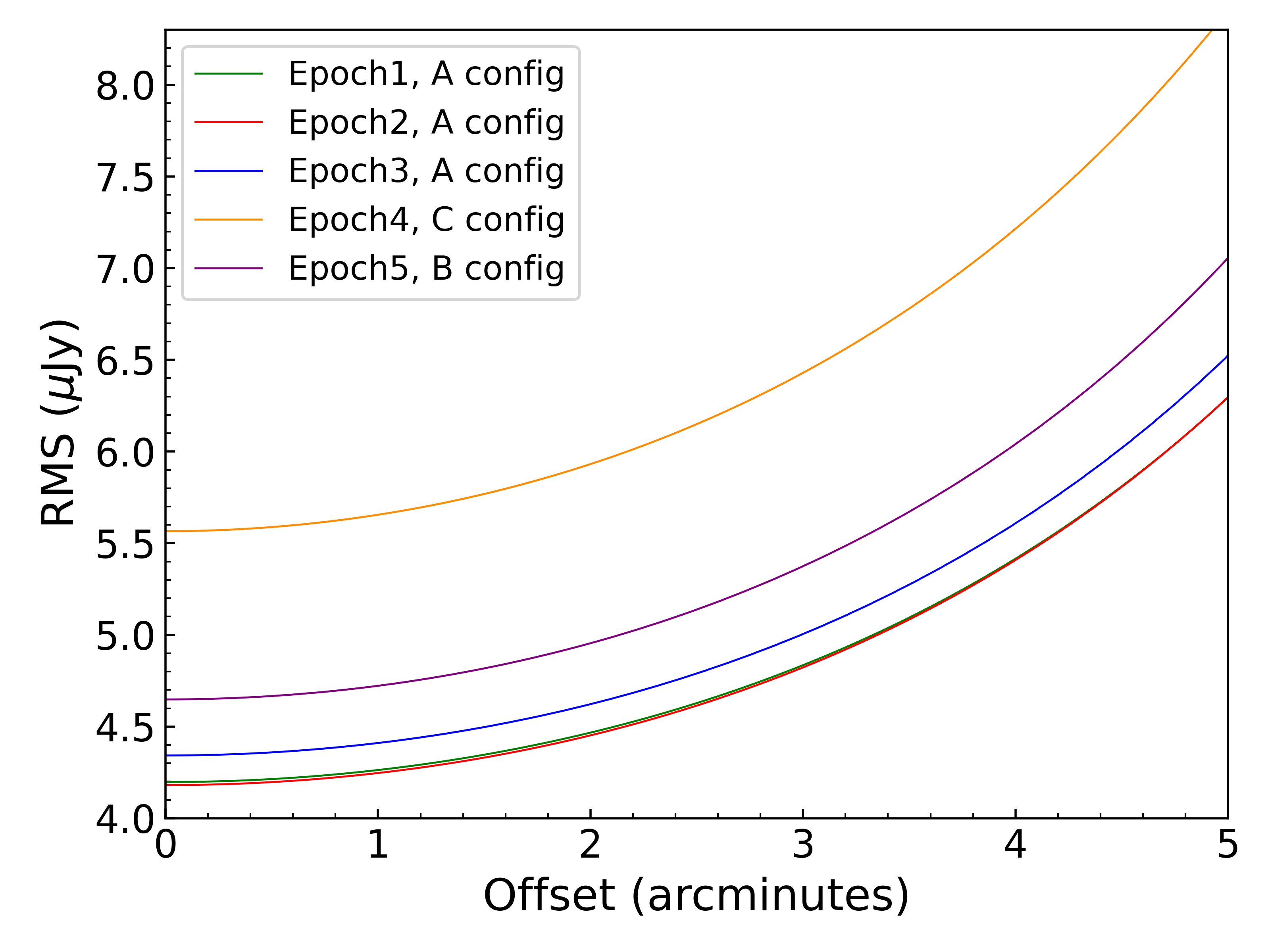}
    \caption{Noise RMS versus offset from the image center for the VLA observations (over the five epochs E1-E5) of the field of the \textit{Swift}/GUANO candidate counterpart to GW241125 (see Figure \ref{fig:regions}).}
    \label{fig:rms_vals}
\end{figure}

Over the first three epochs, we reached a typical central image root-mean-square (RMS) noise of $\approx 4.2\,\mu$Jy. The RMS noise for each epoch was estimated with \texttt{imstat} using a circular region of radius $10\times$ the nominal synthesized beam width specified in Table \ref{tab:epochs}. The RMS sensitivities of our images across the 5\,$\arcmin$-radius sky region of interest for this analysis are shown in Figure\,\ref{fig:rms_vals}. 

We visually inspected the calibrated images and identified sources with signal-to-noise ratios (SNR) greater than $\approx 7$ in any of the first three VLA epochs (those that have the best resolution; see Table  \ref{tab:epochs}) by dividing the \texttt{imstat} peak flux by the RMS noise at each source's location (see Figure \ref{fig:rms_vals}). We report all of these 25 sources in Table \ref{tab:sourcesall2}.  
For sources with SNR$>10$, coordinates were calculated with the \texttt{imfit} algorithm, using an initial circular region of radius comparable to the FWHM of the synthesized beam, centered around the source position as determined through visual inspection. The \texttt{imfit} algorithm does not perform well on sources with SNR$<10$, so those positions were determined using the location of the peak flux from the \texttt{imstat} algorithm. The position errors were calculated by dividing the semi-major axis of the synthesized beam by the source SNR and adding a 0.1\,$\arcsec$ systematic position error in quadrature. We then used \texttt{imstat} to determine the peak flux density of each source within a circular region with radius determined so as to enclose each source, based on the source size as returned by the initial \texttt{imfit} results. If no \texttt{imfit} measurement is available, we use a circular region with radius equal to the nominal FWHM of the synthesized beam (see Table \ref{tab:epochs}). As a check for potential extended emission in sources with SNR$> 10$, we compared the peak flux density and the integrated flux density as returned by the \texttt{imfit} algorithm (these should be in agreement within measurement errors for point sources). In Table \ref{tab:sourcesall2}, the peak flux density errors are calculated by adding the primary-beam-corrected RMS noise at the location of each source (Figure \ref{fig:rms_vals}) and a nominal 5\% absolute flux calibration error in quadrature.

\begin{figure}
    \includegraphics[width=1\linewidth]{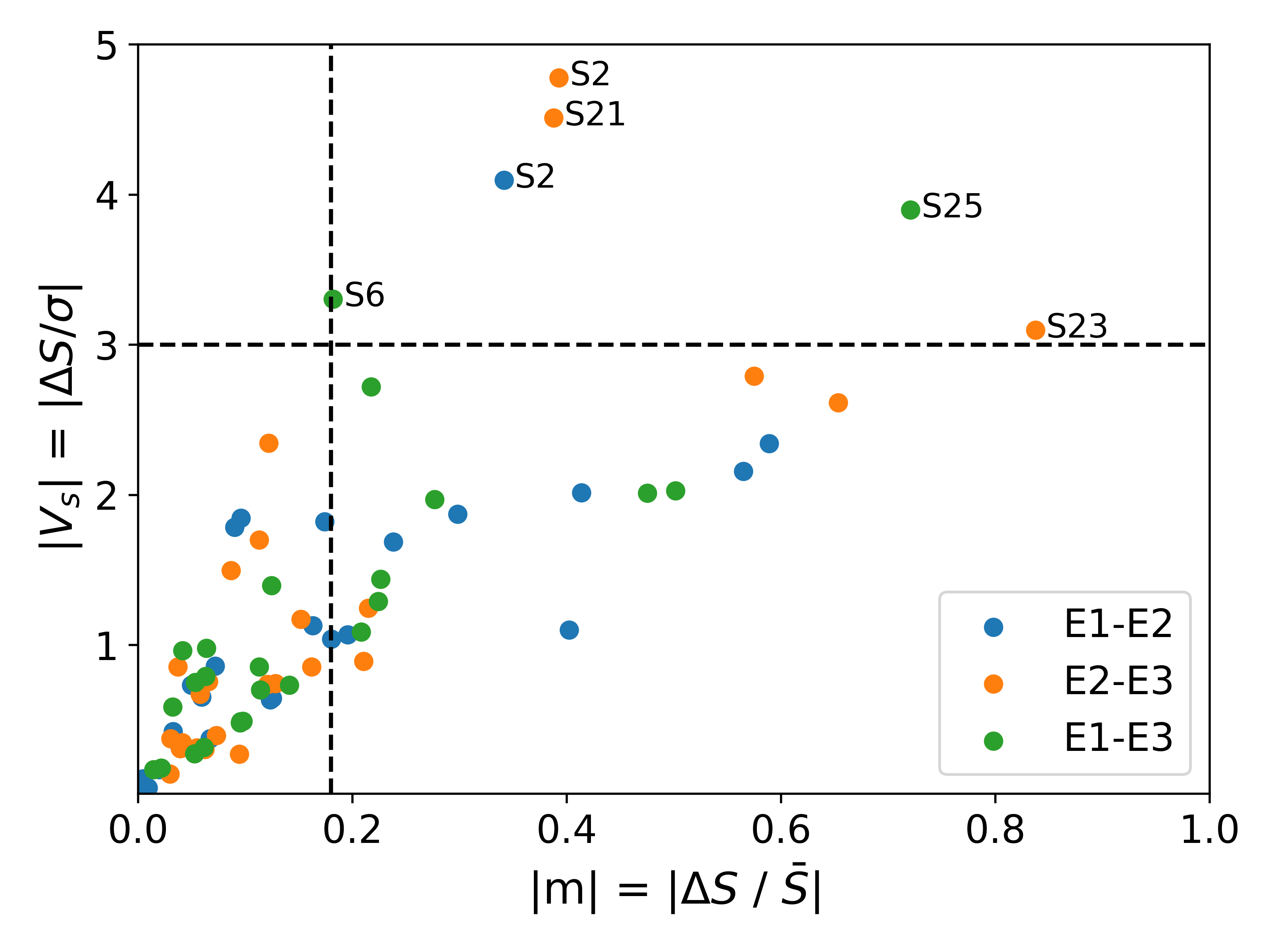}
    \caption{The variability statistic ($V_s$) versus the modulation index ($m$) for the 25 radio sources (Table \ref{tab:sourcesall2}) in VLA images of the localization region of the \textit{Swift}/GUANO candidate counterpart GW241125n  (see Section \ref{sec:obs} for details). The colors indicate the epoch pair while the dashed lines show the selection criteria for variability, those falling above and to the right are variable sources (see Table \ref{tab:vary}).}
    \label{fig:variability}
\end{figure}

\begin{figure}
    \centering
    \includegraphics[width=1\linewidth]{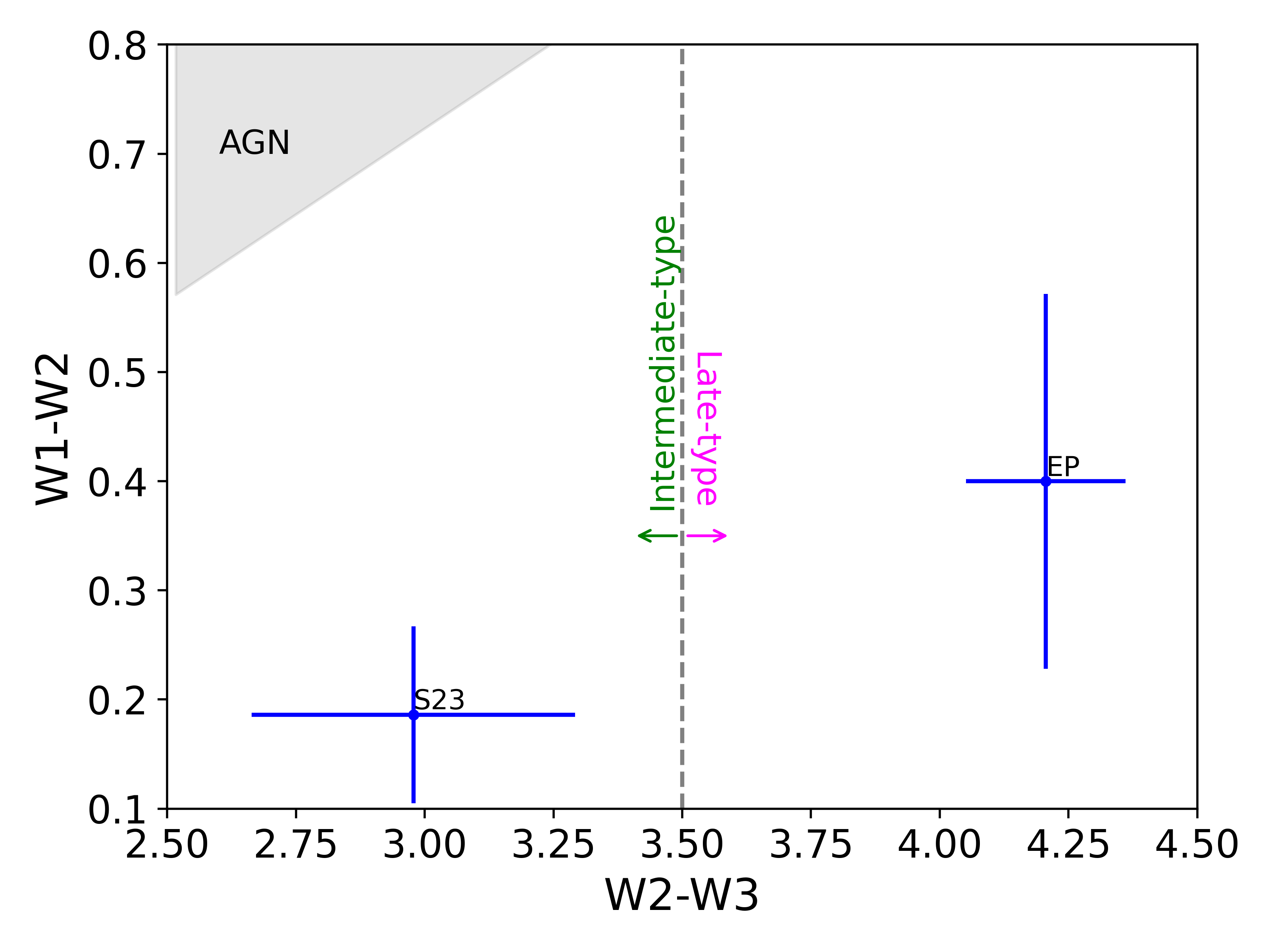}
    \caption{WISE color–color diagrams of the radio sources listed in Tables \ref{tab:vary} and \ref{tab:epsource} that are associated with a known source with WISE colors available. The gray AGN region ($W2-W3 > 2.517$, $W1-W2 >0.315 \times (W2-W3) - 0.222)$ is from \citet{AGNWiseWedge}. The $W2-W3$ color can be used to track galaxy  morphology and star formation rate \citep{Jarrett_2017}, with $2<W2-W3<3.5$ for intermediate-type (disk) galaxies, and $W2-W3>3.5$ for late-type disk galaxies (star-forming disks). }
    \label{fig:wise}
\end{figure}

For all of the sources listed in Table \ref{tab:sourcesall2}, we calculated the variability statistic following \cite{VariabilityStat}, $V_s = \Delta S / \sigma$ where $\Delta S$ is the difference between the flux in each pair of epochs (E1-E2, E1-E3, E2-E3) and $\sigma$ is calculated as the images RMS noise values added in quadrature. We also calculated the modulation index $m = \Delta S / \bar{S}$ where $\bar{S}$ is the average of the flux between epochs. Following \cite{VariabilityStat}, we identify as variable radio sources those with $|V_s| > 3$ and $|m| > 0.18$, as shown in Figure \ref{fig:variability}. For the sources showing this variability, we utilized the VLA image taken with the array in its more compact configurations (E4 and E5) to check for any evidence for extended emission. Following \citet{ChandraDeep}, we consider a source to be point-like if the following two criteria are met: (i) the ratio of its integrated flux density (as estimated using \texttt{imfit}) and its peak flux density is in the 0.9--1.5 range; (ii)  the source major (minor) axis (as estimated using \texttt{imfit}) is less than $1.5\times$ the clean beam major (minor) axis. Some sources (those marked with ``M'' in Table \ref{tab:vary}) show signs of being marginally resolved in one or two epochs: their flux ratio falls outside the point source range in (i), but \texttt{imfit} is not able to report a size measurement due to the lower SNR. None of the variable radio sources match both the above criteria for being classified as extended.  Finally, for all our variable sources found in our analysis, we searched the NASA/IPAC Extragalactic Database (NED\footnote{\url{https://science.nasa.gov/astrophysics/data/nasa-ipac-extragalactic-database-ned/}}) for the nearest known source (and record the corresponding angular offset in Table \ref{tab:vary}). For these  NED sources, we used VIZIER to obtain any available WISE colors or redshifts. WISE colors can be useful to identify a potential AGN origin of their emission \citep{AGNWiseWedge}. Only sources S2 and S23 have WISE colors available. Source S2 is consistent with a known galaxy at $z=0.24$ as reported by the REGALADE galaxy catalog \citep{RegaladeCat} which is much closer than the GW-derived range for GW241125 (Section \ref{sec:detection}).  Hence, we do not include S2 in Figure \ref{fig:wise}. 

\begin{deluxetable*}{cccccccc}[ht]

\tablecolumns{8}
\tablewidth{1.0\columnwidth} 
\tablecaption{Radio sources identified in the VLA images of the field of the \textit{Swift}/GUANO candidate counterpart to GW241125. Here we only report the sources whose flux density fulfills our selection criteria for variability. } 
\tablehead{ \colhead{\textbf{ID}} & \colhead{\textbf{Epoch}} & \colhead{\textbf{Source}} & \colhead{\textbf{R.A. Dec.}} & \colhead{\textbf{Pos.Err}} & \colhead{\textbf{Peak F$_\nu$}} & \colhead{\textbf{Closest Known Source}} & \colhead{\textbf{Offset}}\\ \colhead{} & \colhead{(days)} & \colhead{\textbf{shape}} & \colhead{(hh:mm:ss deg:mm:ss)} & \colhead{(\arcsec)} & \colhead{($\mu$Jy)} & \colhead{} & \colhead{(\arcsec)}}

\startdata
    S2       & 5   & P & 03:51:58.166 +69:44:51.64 & 0.11 & 106.3 $\pm$ 7.5  & WISEA J035158.19+694451.6 & 0.10\\
    $\cdots$ & 18  & P & $\cdots$                  & 0.12 & 75.3  $\pm$ 6.5 & PS J035158.19+694451.7 & \\
    $\cdots$ & 46  & P & $\cdots$                  & 0.11 & 112.1 $\pm$ 7.9 & $\cdots$ &  \\
    $\cdots$ & 211 & P & $\cdots$                  & 0.28 & 89.4  $\pm$ 8.4 & $\cdots$ & \\
    $\cdots$ & 293 & P & $\cdots$                  & 0.15 & 92.1  $\pm$ 7.5 & $\cdots$ & \\
    \hline
    S6       & 5   & P & 03:51:57.104 +69:42:54.66 & 0.11 & 131.3 $\pm$ 8.1 & WISEA J035157.20+694242.9 & 12 \\
    $\cdots$ & 18  & P & $\cdots$                  & 0.11 & 119.3 $\pm$ 7.7 & $\cdots$ &  \\
    $\cdots$ & 46  & P & $\cdots$                  & 0.11 & 109.4 $\pm$ 7.2 & $\cdots$ &  \\
    $\cdots$ & 211 & P & $\cdots$                  & 0.18 & 131.0 $\pm$ 9.0 & $\cdots$ & \\
    $\cdots$ & 293 & P & $\cdots$                  & 0.13 & 117.5 $\pm$ 7.8 & $\cdots$ &  \\
    \hline
    S21      & 5   & P & 03:52:21.726 +69:37:22.09 & 0.12 & 87.5  $\pm$ 7.0  & 
WISEA J035220.46+693716.9 & 8.3\\
    $\cdots$ & 18  & P & $\cdots$                  & 0.12 & 73.4  $\pm$ 6.6 & unWISE 0586p696o0093160 & 0.30 \\
    $\cdots$ & 46  & P & $\cdots$                  & 0.11 & 108.7 $\pm$ 7.8 & $\cdots$ &  \\
    $\cdots$ & 211 & P & $\cdots$                  & 0.23 & 115.4 $\pm$ 9.3 & $\cdots$ &  \\
    $\cdots$ & 293 & P & $\cdots$                  & 0.15 & 103.6 $\pm$ 8.0 & $\cdots$ &  \\
    \hline 
    S23      & 5   & P & 03:52:23.753 +69:43:51.21 & 0.22 & 21.5 $\pm$ 4.7  & 
WISEA J035223.78+694351.3 & 0.20\\
    $\cdots$ & 18  & M  & $\cdots$                  & 0.33 & 14.3 $\pm$ 4.7 & $\cdots$ &\\
    $\cdots$ & 46  & P & $\cdots$                  & 0.15 & 34.9 $\pm$ 5.1 & $\cdots$ & \\
    $\cdots$ & 211 & P & $\cdots$                  & 0.83 & 24.1 $\pm$ 6.2 & $\cdots$ &  \\
    $\cdots$ & 293 & P & $\cdots$                  & 0.32 & 31.2 $\pm$ 5.4 & $\cdots$ &  \\
    \hline
    S25      & 5   & P & 03:52:54.129 +69:39:20.50 & 0.13 & 55.1 $\pm$ 5.9& 
WISEA J035256.71+693913.0 & 15 \\
    $\cdots$ & 18  & P & $\cdots$                  & 0.15 & 46.8 $\pm$ 5.7  & unWISE 0586p696o0093847	& 0.20\\
    $\cdots$ & 46  & M  & $\cdots$                  & 0.20 & 25.9 $\pm$ 5.5 & $\cdots$ & \\
    $\cdots$ & 211 & P & $\cdots$                  & 0.42 & 54.3 $\pm$ 7.4 & $\cdots$ &  \\
    $\cdots$ & 293 & P & $\cdots$                  & 0.26 & 44.5 $\pm$ 6.2 & $\cdots$ &  \\
\enddata 
\tablecomments{
Columns in this Table are: the source name specified in Table \ref{tab:sourcesall2}, epoch of observation (time since GW BBH merger), source shape (P: point-like; M: marginally resolved; R: resolved), the R.A, Dec. position as measured in epoch 1, position error on source in each epoch, peak flux of source in each epoch, nearest known source identifier, and angular separation to the nearest known source match from NED.}
\label{tab:vary}
\end{deluxetable*}

\begin{figure}
    \centering
    \includegraphics[width=1\linewidth]{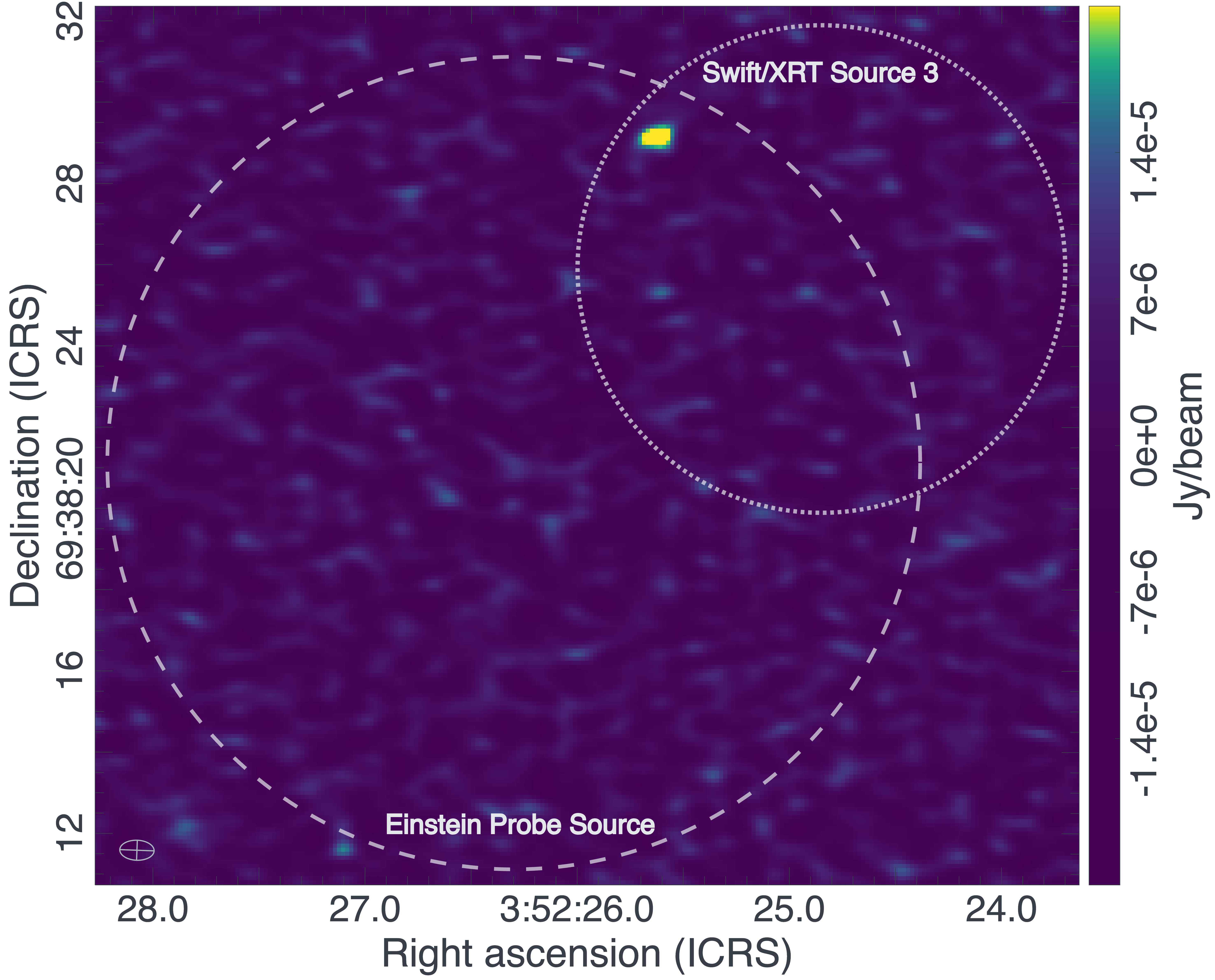}
    \caption{Epoch 1 VLA image of a portion of the GW241125 field, showing the EP source region (10\arcsec radius; larger dashed circle) and the \textit{Swift}/XRT source X3 region (4.7\arcsec radius; smaller dotted circle). The radio source at the top of the intersection has an SNR of $\approx 6.5$ in this epoch.}
    \label{fig:epimage}
\end{figure}

In addition to this general search for variable radio sources in the \textit{Swift}/GUANO field, we performed a targeted search looking specifically at X-ray sources identified in the \textit{Swift}/GUANO region by the \textit{Swift}/XRT (Section \ref{sec:xrt}) and the EP (Section \ref{sec:detection}). One \textit{Swift}/XRT source, X3 that we describe in the next Section, was identified within the GUANO uncertainty region (Figure \ref{fig:regions}).  Within the 10$\arcsec$ EP uncertainty region (Figure \ref{fig:epimage}, dashed line), we found one radio source that also overlaps the 4.7$\arcsec$ \textit{Swift}/XRT X3 uncertainty region (Figure \ref{fig:epimage}, dotted line). We report the radio flux densities measured for this source across our VLA observations in Table\,\ref{tab:epsource}. Radio source X3 does not show any significant variability across our five VLA epochs, and its SNR remains at $\approx 6$ throughout. The nearest source in NED is a WISE source located $\approx 0.66\arcsec$ away, dubbed WISEA J035225.57+693829.7 \citep{WISE1}. The WISE colors of this source are shown in Figure \ref{fig:wise}. 

\begin{deluxetable*}{cccccccc}[]

\tablecolumns{8}
\tablewidth{1.0\columnwidth} 
\caption{Properties of the radio counterpart of the EP source identified in the error region of the \textit{Swift}/GUANO candidate counterpart to GW241125.} 
\tablehead{ \colhead{\textbf{Source}} & \colhead{\textbf{Epoch}} & \colhead{\textbf{R.A. Dec.}} & \colhead{\textbf{Pos. Err}} & \colhead{\textbf{Peak F$_x$}} & \colhead{\textbf{E1-Offset}} & \colhead{\textbf{Closest Known Source}} & \colhead{\textbf{Offset}}\\ \colhead{} & \colhead{(days)} & \colhead{(hh:mm:ss deg:mm:ss)} & \colhead{(\arcsec)} & \colhead{($\mu$Jy)} & \colhead{(\arcsec)} & \colhead{} & \colhead{(\arcsec)}}
\startdata
    EP/X3 & 5   & 03:52:25.630 +69:38:29.107 & 0.17 & 31.3 $\pm$ 5.1 & -  & WISEA J035225.57+693829.7 & 0.66  \\
     & 18  & $\cdots$                   & 0.17 & 33.6 $\pm$ 5.1 & 0.10 &  &     \\
     & 46  & $\cdots$                   & 0.17 & 30.9 $\pm$ 5.3 & 0.080 &  &     \\
     & 211 & $\cdots$                   & 0.50 & 42.1 $\pm$ 6.7 & 0.64 &  &    \\
     & 293 & $\cdots$                   & 0.33 & 30.6 $\pm$ 5.6 & 0.25 &  &    \\
\enddata  
\tablecomments{Columns are: source name, days since BBH merger, position, position error, peak flux, offset from epoch 1 position, name of nearest NED source, offset of nearest known NED source. }
\label{tab:epsource}
\end{deluxetable*}

\subsection{Swift/XRT}
\label{sec:xrt}
The X-Ray Telescope (XRT; \citealt{Burrows+2005}) on-board \textit{Swift} began observations of the field of GW241125 at $\sim$ 15 hours after the BAT-GUANO trigger \citep{GCN38324}. In total, 30.4\,ks of exposure were obtained in 7 individual epochs between 25 November 2024 and 1 January 2025 UTC. 

\begin{figure}
    \includegraphics[width=1\linewidth]{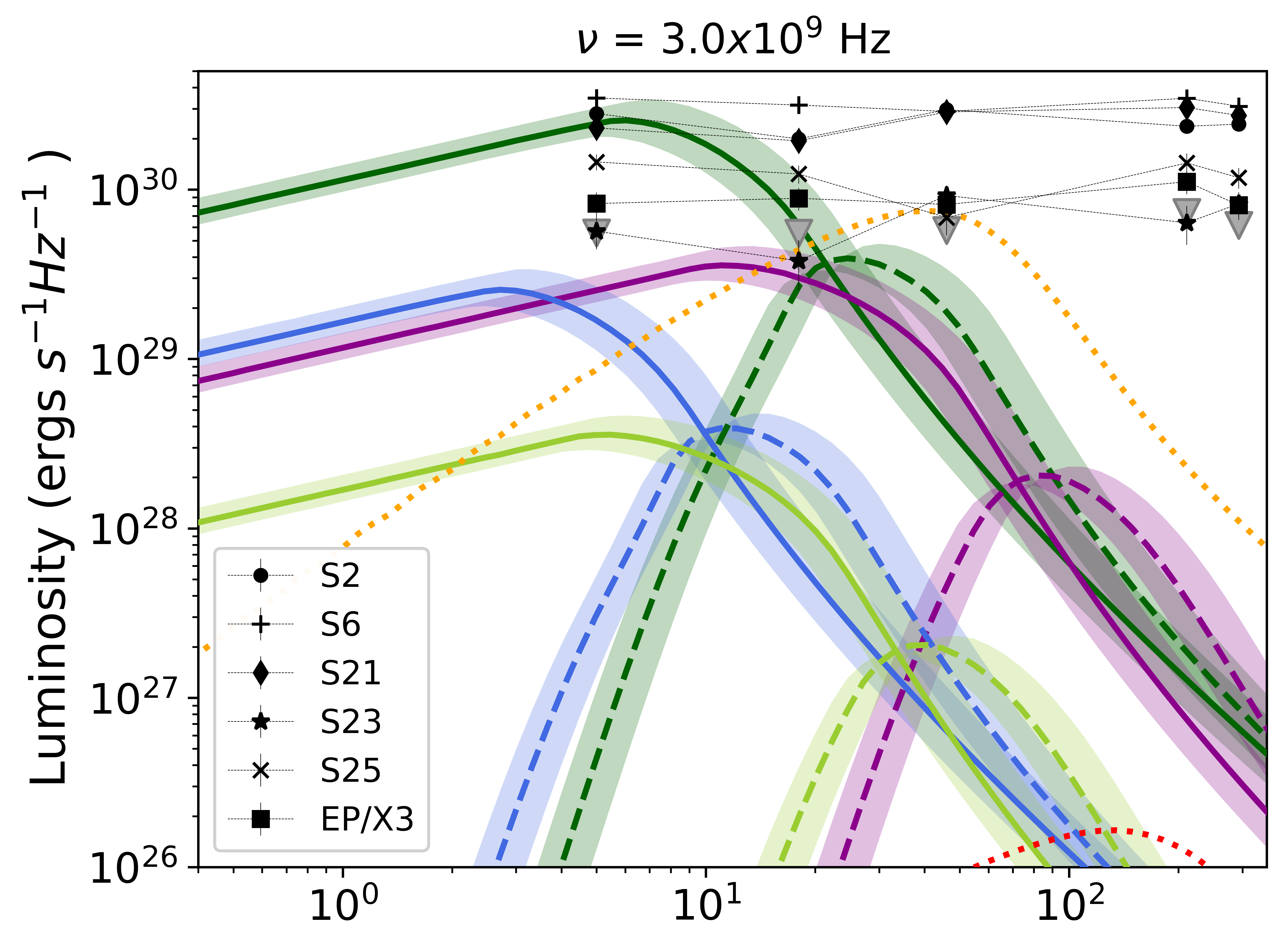}
    \includegraphics[width=1\linewidth]{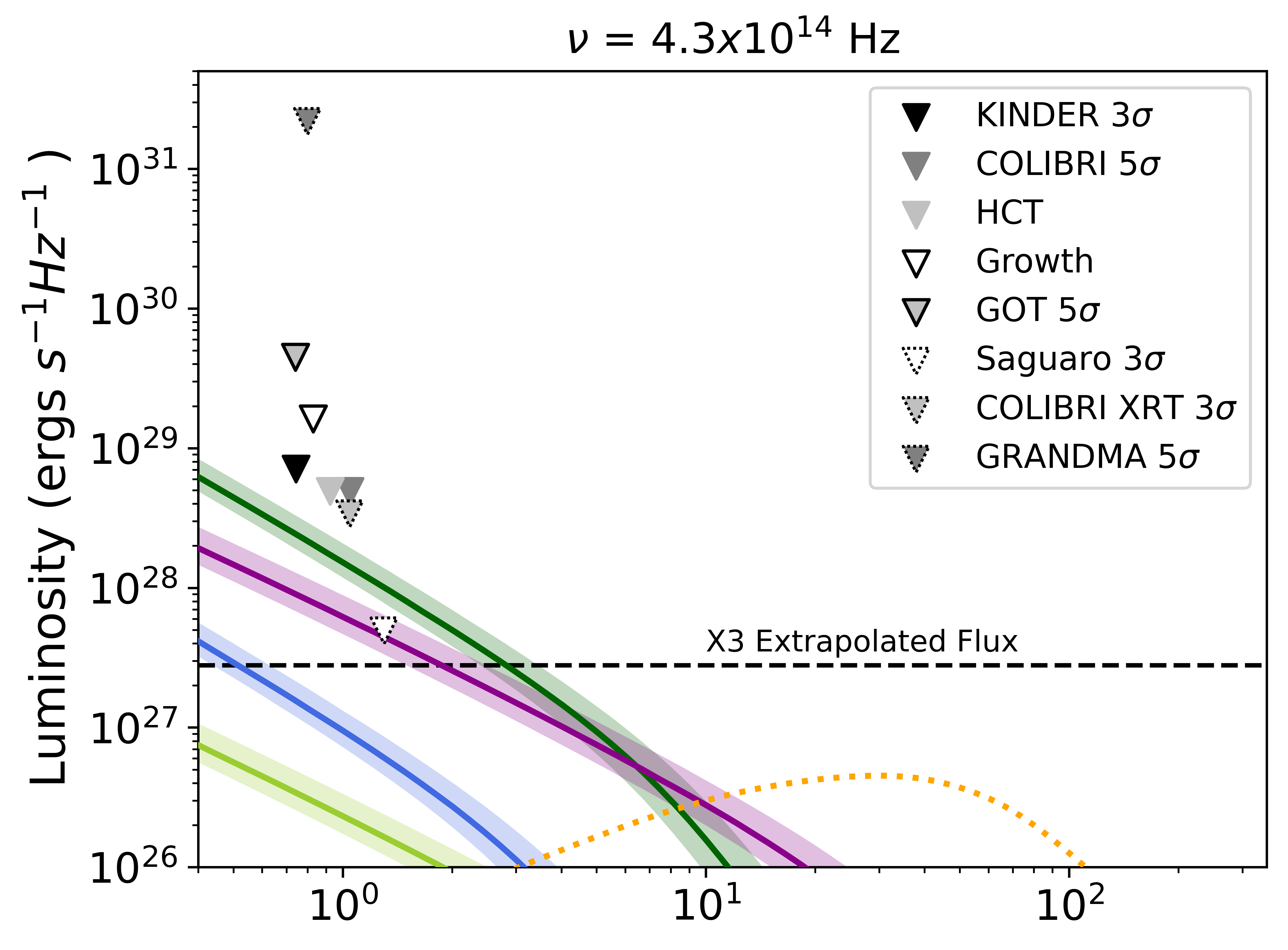}
    \includegraphics[width=1\linewidth]{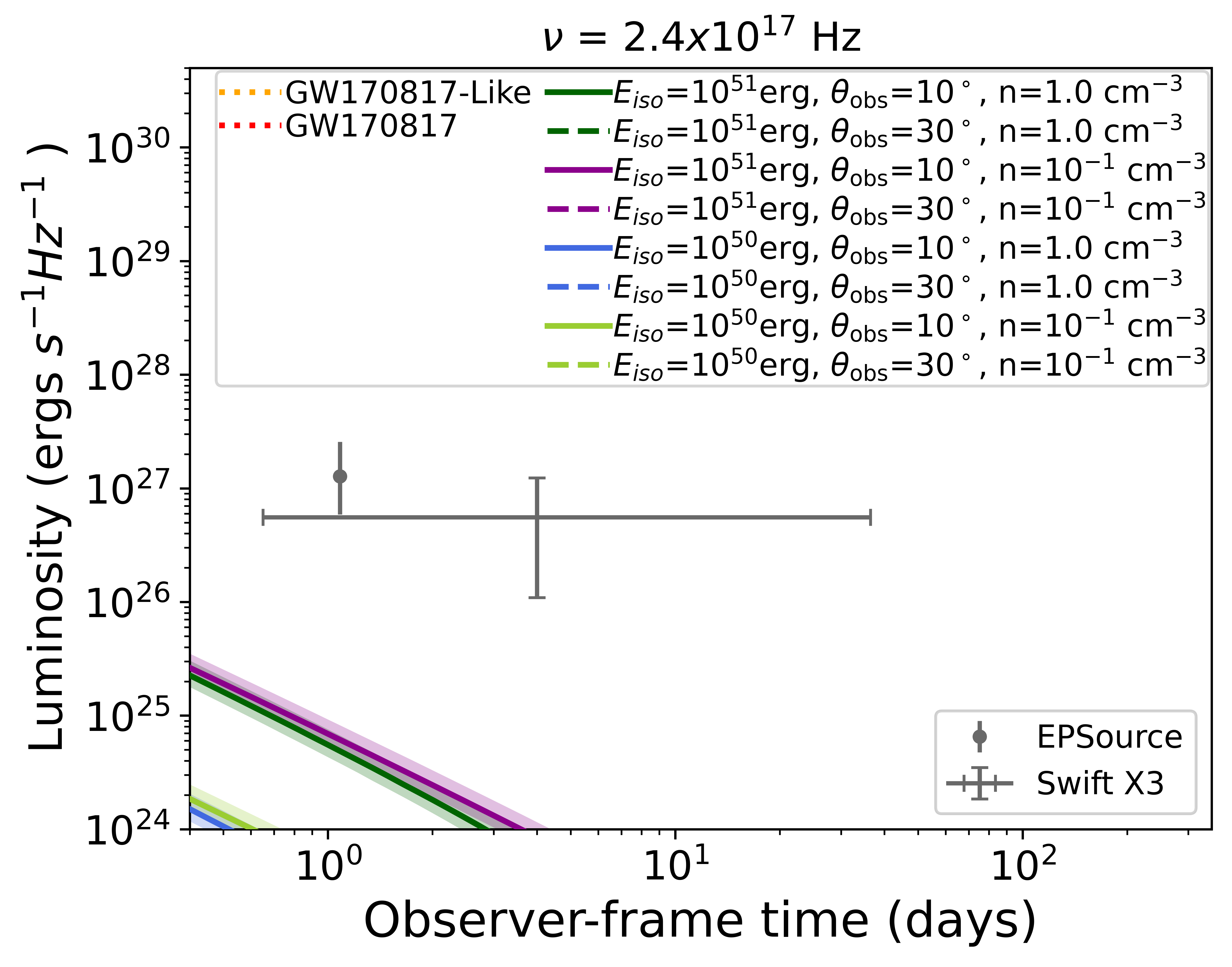}
    \caption{\texttt{afterglowpy} light curves for top-hat jets with $\theta_{jet}=10$\,deg (lines). Observing angles $\theta_{\rm obs}$, isotropic-equivalent kinetic energy and ISM density values are given in the legend (bottom panel). We assume $d_L \approx 4.7$\,Gpc (shaded regions for the distance uncertainty), $p=2.5$, and $\varepsilon_e=\epsilon_B=0.1$. Red curves show GW170817 and a GW170817-like afterglow with enhanced ISM density.  TOP: Gray triangles mark VLA $5\sigma$ limits at epochs E1–E5, and black symbols show variable field sources (Table \ref{tab:vary}) as well as EP/X3. CENTER: Symbols mark optical upper limits from follow-up telescopes (see legend). BOTTOM: Symbols mark the candidate X-ray counterpart EP/X3  (see text).}
    \label{fig:models1}
\end{figure}

Consistent with the findings reported in \citet{GCN38324}, only a single source is confidently detected inside the BAT-GUANO error circle (referred to as S241125n-X3), both in the stacked frame and in individual epochs. In the stacked frame we find a position for this source of $\alpha$=03$^{\mathrm{h}}$52$^{\mathrm{m}}$25.64$^{\mathrm{s}}$, $\delta$=+69$^{\circ}$38\arcmin28.7\arcsec, with a 90\% confidence positional uncertainty of 4.7\arcsec. S241125n-X3 does not show any evidence for variability in the 7 epochs (between 0.6 and 36 days since the GW trigger) of XRT observations, with an average count rate of $1.20^{+0.27}_{-0.24} \times 10^{-3}$\,s$^{-1}$. For an assumed power-law spectrum with a photon index of $\Gamma$=1.7 and foreground Galactic extinction $N_{H} = 3 \times 10^{20}$\,cm$^{-2}$, this corresponds to a 0.3--10\,keV flux of $5.1^{+1.1}_{-1.0} \times 10^{-14}$\,erg\,cm\,$^{-2}$\,s$^{-1}$. Assuming a power-law spectrum of the form $F_{\nu} \propto \nu^{-\beta}$, the implied radio-to-X-ray spectral index of X3 is $\beta \approx 0.49$. The extrapolated optical flux density derived assuming this spectral index is shown in Figure  \ref{fig:models1} (horizontal dashed line). 

No X-ray emission is detected at the location of any of the five variable radio sources identified via our VLA observations and reported in Table~\ref{tab:vary}. Adopting the same power-law spectrum used above, the typical 90\% confidence upper limit on the 0.3--10\,keV X-ray flux is $\approx 10^{-14}$\,erg\,cm$^{-2}$\,s$^{-1}$. The X-ray non-detections of these radio sources imply that their radio-to-X-ray spectral index should be steeper than $\beta \gtrsim 0.56-0.65$ (where for each source we have assumed a radio flux density equal to their average radio flux density across E1-E5). 

\section{Discussion}
\label{sec:discussion}

Our radio follow-up observations of the field of the \textit{Swift}/GUANO candidate $\gamma$-ray counterpart to GW241125 has identified several variable radio sources compatible with the GUANO localization, but none of these sources is compatible with the faster transient-like evolution that would be expected in the mini-disk BBH scenario. This is shown in more detail in Figure\,\ref{fig:models1}, where 
we compare model GRB afterglow light curves obtained using \texttt{afterglowpy} \citep{AfterglowPy} with radio-to-X-ray observations of the GUANO field. Because the environment around BBHs is expected to be free of the neutron-rich debris that is expected to characterize BNS merger sites and that has been linked to the emergence of structured relativistic jets (as opposed to top-hat ones), here we work under the hypothesis that a simple top-hat jet is powered by the accretion of the mini-disk around the GW241125 BBH.  

For the top-hat jet afterglow models, we assume a jet opening angle of $\theta_{\rm jet}=10$\,deg \citep{sGRBs, GhirlandaGRBs, FongGRBs}, and vary the isotropic energy $E_{\rm iso}$ within $10^{49} - 10^{51}$ erg, consistent with typical isotropic energies of short GRBs \citep{sGRBs} and the high end of energies considered in \cite{EM_BBH_Limits} which were centered on the estimated $E_{iso}$ for GW150914 of $\approx10^{49}$\,erg. We also vary the density $n_{\rm ISM}$ of the interstellar medium (ISM) in which the jet expands within the $0.01-1$\,cm$^{-3}$ range, chosen to represent the environments that may be found in host galaxies of BBH mergers around the merger sites \citep[see e.g., ][for discussion]{PernaHorizons}. 

In the fireball model of GRB afterglows \citep[e.g.,][]{Sari98, PiranRev, GRRev}, the microphysics of the jet is encoded in the parameters $\varepsilon_e$ and $\varepsilon_B$, that quantify the fraction of energy going into accelerating electrons and amplifying magnetic fields behind the shock front, respectively. Because these parameters can span several orders of magnitudes and are hard to constrain without extensive multi-band monitoring, here we fix them to representative values of $\varepsilon_e=\varepsilon_B=0.1$ \citep[e.g.,][]{Sari98, GRRev}.  For a direct comparison with what could be expected in the case of a binary neutron star merger, in Figure \ref{fig:models1}  we also plot the model light curve of a GW170817-like jet. Specifically, the red curve in this Figure is a model that follows closely the analysis of \cite{GW170817Afterglow}, namely, a Gaussian jet with $E_{iso}$ = $10^{53}$ ergs, $\theta_{obs}$ = 0.54 rad, $\theta_{core}$= 0.088 rad, $\theta_{wing}$ = 0.6 rad, $n_{ISM}$ = 0.02 cm$^{-3}$, $\varepsilon_e$ = 0.01, and $\varepsilon_B$ = 0.0002. The orange lines in Figure \ref{fig:models1} are for a GW170817-like jet with the same parameters as above except for $\varepsilon_e$, $\varepsilon_B$, and the the ISM density which in this case are set to $\varepsilon_e$ = 0.08, $\varepsilon_B$ = 0.01, and $n_{\rm ISM}$= 1.0 cm$^{-3}$. 

As evident from Figure\,\ref{fig:models1}, none of the jet afterglow models considered here can account for the behavior of the variable radio sources identified in our search, all of which show radio emission persisting even in the late epochs at more than 100\,days since the GW trigger. 

\begin{figure}
    \centering
    \includegraphics[width=1.0\linewidth]{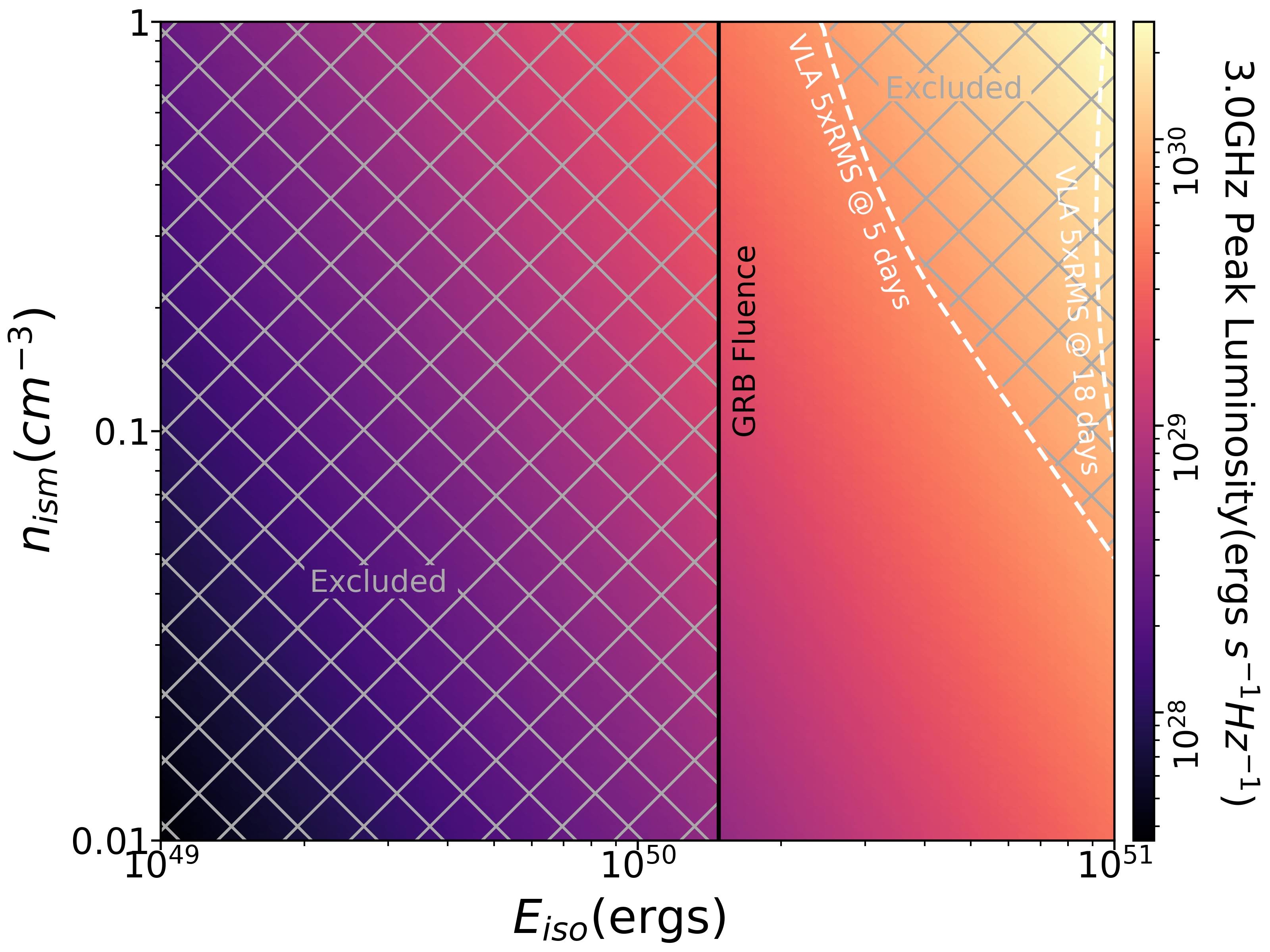}
    \caption{Constraints set on the mini-disk model for BBHs. The color scale measures the peak radio (3\,GHz) afterglow luminosity of a top-hat jet observed on edge ($\theta_{\rm obs}=10$\,deg) as a function of $n_{\rm ISM}$ and $E_{\rm iso}$ (we leave the other model parameters unchanged; see Section \ref{sec:discussion}).  The vertical line indicates the isotropic equivalent energy for a GRB at the distance of GW241125 with $\gamma$-ray fluence as measured by the \textit{Swift}/GUANO for the candidate GRB counterpart to GW241125. The dashed line marks the constraints derived using our VLA observations of the GUANO field, namely, events with $n_{\rm ISM}$ and $E_{\rm iso}$ in the region to the right of the dashed line would have a radio luminosity density at 5\,days since the GW trigger (E1) in excess to the $5\sigma$ sensitivity of our 3\,GHz VLA observations. A similar contour is also plotted for VLA constraints obtained at 18 days since the GW trigger (E2).}
    \label{fig:exclusions}
\end{figure}

In Figure \ref{fig:exclusions} we show the range of energies and ISM densities that are compatible with constraints set by our VLA observations for a top-hat jet observed edge-on ($\theta_{\rm obs}=10$\,deg). We also show in this plot the constraint set by the requirement that the kinetic energy powering the jet as estimated from the afterglow emission is larger than the energy observed in $\gamma$-rays as estimated from the \textit{Swift}/GUANO trigger. While modeling of the $\gamma$-ray emission is beyond the scope of this analysis, we note that in the case of a top-hat jet, an edge-on observer could still see the $\gamma$-ray signal, while explaining the origin of the $\gamma$-rays for an observer located farther off-axis would require invoking a structured jet or an origin for the $\gamma$-rays other than synchrotron emission powered by internal shocks in the jet \citep[e.g., photospheric emission, cocoon breakout, or external shock origin; see ][and references therein for discussion]{2020FrASS...7...78L}. As evident from Figure \ref{fig:exclusions}, the joint GUANO-VLA observations constrain the isotropic-equivalent kinetic energy of a putative jet associated with the \textit{Swift}/GUANO trigger to $\lesssim 3\times10^{50}$\,erg for an $n_{ISM} \approx$ 1.0 cm$^{-3}$. 

Based on the above analysis, we conclude that it is unlikely that the variable radio sources listed in Table \ref{tab:vary} or the source \textit{Swift}/XRT X3 could represent GRB jet afterglows of the candidate GUANO event found in the error region of GW241125. It is also unlikely that a top-hat GRB-like jet with parameters similar to the regions excluded in Figure \ref{fig:exclusions} was launched in the GW241125 merger. However, several regions of the $E_{\rm iso}-n_{\rm ISM}$ parameter space remain possible and are unconstrained by our analysis.

In light of the above, in the remainder of this Section we turn our attention to what could be the origin of the variable radio sources found in our analysis. The most interesting radio source in our sample is perhaps the one found to be spatially coincident with the EP and the \textit{Swift}/XRT source X3. The location of this source in the WISE color diagram in Figure \ref{fig:wise} suggests a star-formation origin of the radio emission. More specifically, as highlighted by \citet{Jarrett_2017}, $W2-W3$ color can be used to track galaxy  morphology and star formation rate, with $W2-W3 < 2$  for ellipticals, $2<W2-W3<3.5$ for intermediate-type (disk) galaxies, and $W2-W3>3.5$ for late-type disk galaxies (star-forming disks). The X3 colors do not suggest an AGN origin as the $W1-W2$ color falls outside the AGN wedge \citet{AGNWiseWedge} and well below the simple one-color cut of $W1-W2>0.8$ suggested by \citet{Stern_2012} for identifying AGN at $z\lesssim 1$. On the other hand, the radio-to-X-ray spectral index of source X3, $\beta \approx 0.49$ (see Section \ref{sec:xrt}), is rather flat, and more in line with that of radio quiet quasars \citep{1994ApJS...95....1E,2006A&A...451...35B} than with the non-thermal spectral indices of $ \gtrsim 0.7$ expected for star-forming galaxies \citep{Tabatabaei_2017}. To check for radio flux potentially missed due to extended emission, we inspected the \texttt{imfit} results for X3, despite its associated radio source not reaching an SNR of 10 in our VLA images. We find no evidence of extended emission in any epoch: the integrated and peak flux densities agree within their measurement errors. 

All of the radio variable sources identified in our VLA observations of the localization area of the \textit{Swift}/GUANO candidate counterpart to GW241125 have X-ray non-detections that constrain their radio-to-X-ray spectral index to be steeper than $\beta \gtrsim 0.56-0.65$. This is compatible with radio emission associated with star formation, as also suggested by the WISE color position of S23, the only radio variable source with $SNR\gtrsim 7$ for which we have WISE colors available. Without redshift information for sources S23 and EP/X3, we cannot securely compare their radio power to that of AGN and star-forming galaxies. However, we can do so assuming they are at the distance of GW241125. The median power for a star forming galaxy in the NRAO VLA Sky Survey (NVSS) is $\approx1.4\times10^{22}$\,W/Hz and for an AGN it is $\approx1\times10^{23}$ W/Hz \citep{NVSS}. At the distance of GW241125, the 1.4\,GHz power for S23 is $\approx1.2\times10^{23}$\,W/Hz and for EP/X3 is $\approx1.8\times10^{23}$\,W/Hz (these estimates assume a radio spectral index of 1). These values are more consistent with an AGN, but still within the range of star-forming galaxies whose maximum value is $\approx5.5\times10^{23}$\,W/Hz.

Large radio surveys of the sky have examined the variability of persistent radio sources. The fraction of variable radio sources generally depends on the survey sensitivity limit and, potentially strongly, on the survey frequency. \cite{ChandraDeep} found that the 1.4\,GHz radio sky is relatively quiet at the sub-mJy level, with less $\lesssim 1\%$ of sources being variable. At the higher 3\,GHz radio frequency of our observations, we find that 5 out of 25 sources, or 20\%,show some degree of variability, though only $\approx 10\%$ have $m\ge 0.5$. This is in between the variable fraction of $\lesssim 1\%$ level found in 1.4\,GHz surveys \citep{ChandraDeep} and the $\approx 30\%$ found for studies at 5\,GHz \citep{Ofek_2011}. More recently, \cite{VLASSTransients} used the VLA Sky Survey (VLASS) to study the variability of the radio sky at 3\,GHz, finding $\approx 5\%$ of sources above 20\,mJy are variable.


\begin{figure}[t]
    \centering
    \includegraphics[width=\linewidth]{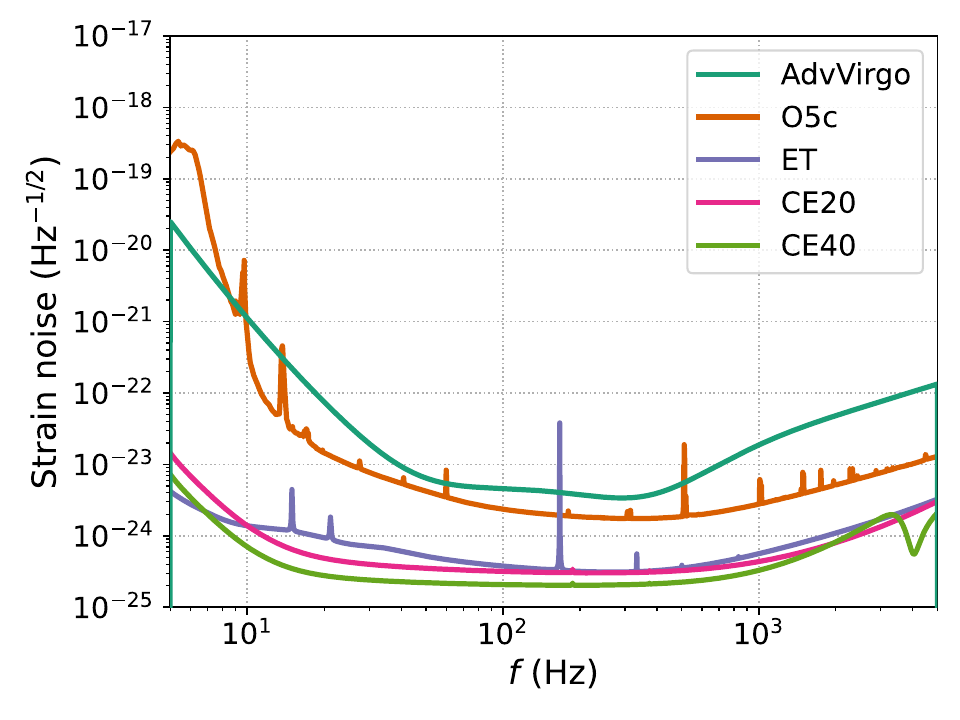}
    \caption{Representative amplitude spectral density curves for current and future ground-based GW detectors considered in this study.}
    \label{fig:psds}
\end{figure}

\section{Prospects for probing the BBH mini-disk scenario}
\label{sec:nextgen}

In what follows, we place the results of the radio follow-up campaign of GW241125, and more generally the search for radio afterglows for BBHs in the mini-disk model, in the broader context of upcoming and future GW detector runs. 

To this end, we simulate a one-year population of BBHs  using the Python package \texttt{gwforge}~\citep{Chandra:2024dhf} so as to characterize the distribution of sky localizations and luminosity distances of the systems that we may detect with current-generation GW detectors of improved sensitivity (teal and orange lines in Figure\,\ref{fig:psds}), as well as with networks of next-generation ground-based GW detectors such as Cosmic Explorer (CE) and the Einstein Telescope (ET) (purple, pink, and green lines in Figure\,\ref{fig:psds}). More specifically, we consider two detector networks:
\begin{itemize}
    \item An O5 network with the two LIGO detectors at O5c/A+ sensitivity~\citep{O5c_sensitivity} and Virgo at Advanced Virgo design sensitivity~\citep{VIRGO:2014yos}. 
    \item A next-generation network with a 40\,km CE (CE40) at default sensitivity in \cite{CE_sensitivity}, a 20 km CE (CE20) at default sensitivity for CE20 in \cite{CE_sensitivity}, and an ET in triangle configuration at ET10 sensitivity~\citep{ET_sensitivity_CoBA,ET:2025xjr}.
\end{itemize}

To generate our synthetic population, we adopt the \textsc{Power-Law + Peak} model described in \cite{KAGRA:2021duu}. The component masses ($m_i$) are distributed within the range [2, 100]~$M_\odot$, while the mass ratio ($q = m_1/m_2$) spans the range [1, 50]. 
The binary spins are sampled assuming an isotropic spin distribution. The probability density functions for the spin magnitudes ($a_i$) and spin angles ($\theta_i$, $\phi_{12}$, $\phi_\text{JL}$) are listed in Table \ref{table:pop_models_extrinsic}.\footnote{Additional details on the spin parameters \texttt{a\_1, a\_2, tilt\_1, tilt\_2, phi\_12}, and \texttt{phi\_jl} are available in \cite{pesummary}.} The table also summarizes the distributions and parameter ranges adopted for the inclination angle ($\theta_\text{JN}$), sky-location angles ($\alpha, \delta, \Psi$), and coalescence phase ($\phi_c$) used in the binary simulations.
\begin{table}
\centering
\def\arraystretch{1.3}
\begin{tabular}{ccc}
\hline
\hline
\textbf{Parameter} & \textbf{Prior} & \textbf{Range} \\ \hline
$m_1$, $m_2$ & \textsc{Power-Law + Peak} & $2 \text{ - } 100$ \\ \hline
$q=m_1/m_2$ & \textsc{Power-Law + Peak} & $1 \text{ - } 50$ \\ \hline
$z$ & \begin{tabular}[c]{@{}c@{}}\textsc{Madau-Dickinson}\\ with time-delay\end{tabular} & $0 \text{ - } 10$ \\ \hline
$a_1$, $a_2$ & Uniform & $0 \text{ - } 0.99$ \\ \hline
$\theta_1$, $\theta_2$ & Uniform sine & $0 \text{ - } \pi$ \\ \hline
$\phi_{12}$ & Uniform & $0 \text{ - } 2\pi$ \\ \hline
$\phi_\text{JL}$ & Uniform & $0 \text{ - } 2\pi$ \\ \hline
$\theta_\text{JN}$ & Uniform sine & $0 \text{ - } \pi$ \\ \hline
$\phi_c$ & Uniform & $0 \text{ - } 2\pi$ \\ \hline
$\alpha$ & Uniform & $0 \text{ - } 2\pi$ \\ \hline
$\delta$ & Uniform cos & $-\pi/2 \text{ - } \pi/2$ \\ \hline
$\psi$ & Uniform & $0 \text{ - } \pi$ \\ \hline
\end{tabular}
\caption{Probability distribution models and ranges for parameters used for simulating binaries.}
\label{table:pop_models_extrinsic}
\end{table}

To obtain a redshift distribution representative of the population accessible to next-generation detector networks, we employ the Madau-Dickinson star formation rate model~\citep{Madau:2014bja} together with an inverse time-delay distribution in redshift~\citep{Dominik:2012kk, Dominik:2013tma, Fishbach:2018edt}.\footnote{See \cite{Divyajyoti:2026gds} for complete details on the redshift model used here and associated parameter values.} We sample redshift in the range $z \in [0, 10]$ and adopt a local merger rate of $22$\,Gpc$^{-3}$\,yr$^{-1}$~\citep{KAGRA:2021duu}.

\begin{figure}
    \includegraphics[width=1\linewidth]{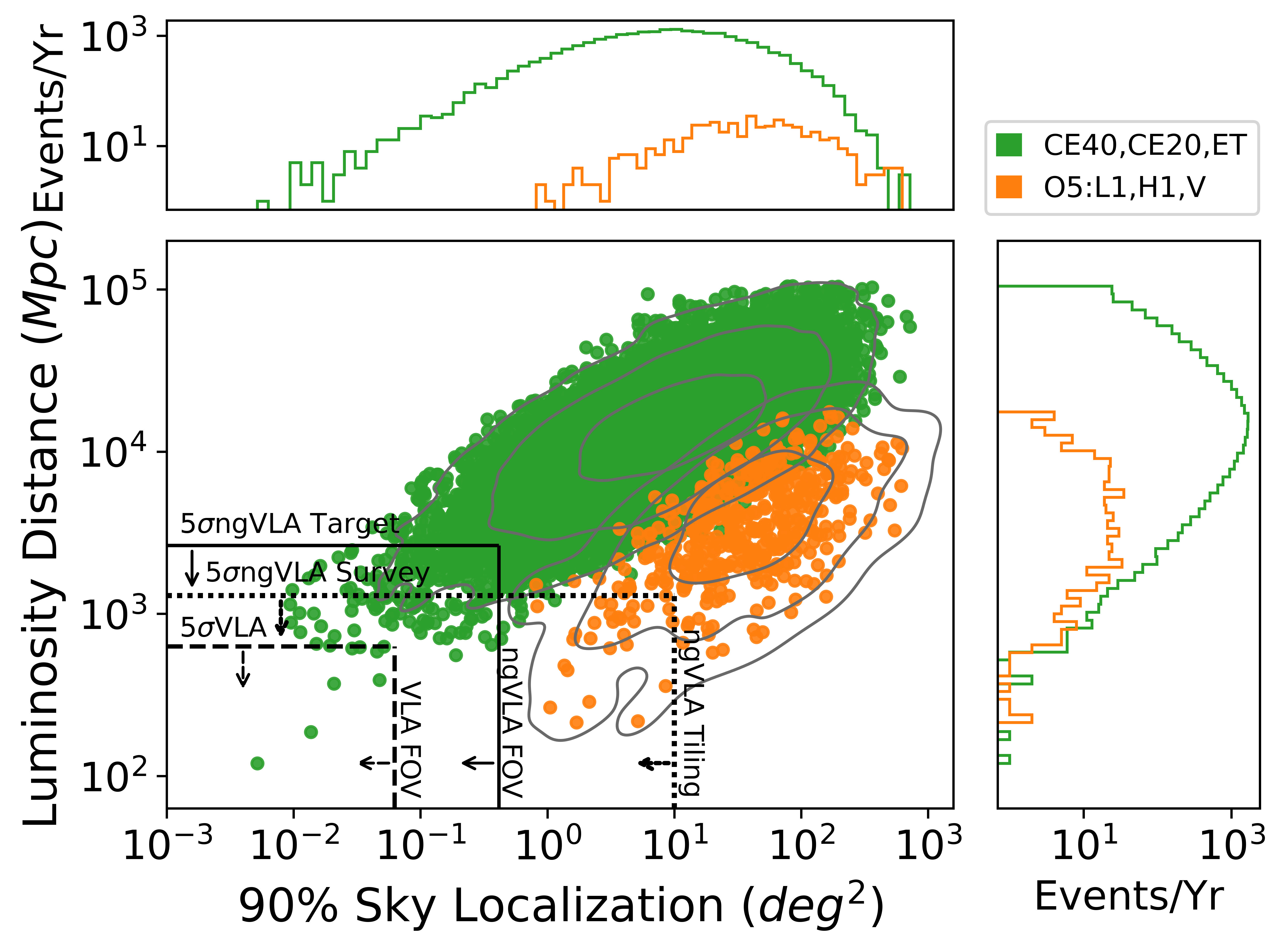}
    \caption{Simulation results for a realistic population of BBHs as detected in 1\,yr with a network of current-generation detectors of improved sensitivity (orange) and of next-generation detectors (green). Contours indicate equal-density regions from a Gaussian kernel density estimate, enclosing 50\%, 90\%, and 99\% of the estimated probability distribution. We mark the areas that are accessible for radio follow up with current-generation radio arrays such as the VLA (dashed lines), and next-generation interferometers such as the next-generation VLA (dotted and solid lines). We assume a radio  luminosity of $10^{28}$ erg\,s$^{-1}$Hz$^{-1}$ to evaluate the distance reach of the VLA/ngVLA  for BBH radio afterglows in the mini-disk scenario. See text for discussion.
    \label{fig:sims}}
\end{figure}
 The results of our simulation are shown in Figure \ref{fig:sims}. For the two detector networks, we plot the BBH GW localization uncertainty against the injected luminosity distance for all simulated BBH events detected with a network signal-to-noise ratio SNR$_{\rm GW}>10$ at Dec.$>-30$\,deg
(so as to be visible from Northern Hemisphere radio facilities), together with the marginal distributions of their detection rates as a function of each quantity. The two GW networks considered here occupy distinct regions of the parameter space. The current-generation network detects BBHs at a rate of order 10 per year, at luminosity distances of $\sim 1-10$\,Gpc and with relatively poor localizations of $\sim 10-100$\,deg$^{2}$ (orange). The next-generation network detects BBHs at a far higher rate, up to $\sim 10^3$ per year,  and extends the reach to cosmological distances, while simultaneously delivering tighter localizations that peak near $ \sim 1-10$\,deg$^{2}$. 

While the next-generation network both localizes BBH events more precisely and detects them in far greater numbers, the bulk of the probed population lies at distances well beyond that of GW241125. Hence, to evaluate the detectability of potential jet radio afterglows in the context of the mini-disk model, we consider the VLA and the next-generation VLA \citep{2018IAUS..336..426M} in three configurations \citep[see also][and references therein]{KaraP}:
\begin{itemize}
\item The VLA in its A configuration operating at 3\,GHz and conducting deep pointed observations, as considered in this study (RMS $\approx 4.2\,\mu$Jy);
\item The ngVLA operating at 2.4\,GHz in its targeted follow up mode (1\,hr RMS of $\approx 0.24\,\mu$Jy);
\item The ngVLA operating at 2.4\,GHz in its so-called survey mode, aimed at tiling 10\,deg$^2$ regions of the sky in 10\,hr with a RMS sensitivity of $\approx 1\,\mu$Jy.
\end{itemize}
The field of view and $5\sigma$ of these three array configurations are shown in Figure \ref{fig:sims}. In this Figure, we assume a radio  luminosity of $10^{28}$ erg\,s$^{-1}$Hz$^{-1}$ to evaluate the distance reach of various radio interferometers for BBH radio afterglows in the mini-disk scenario (see Figure \ref{fig:models1}). 

Without the help of large-field high-energy instruments like the \textit{Swift}/BAT (FOV of $\approx 1.4$\,sr or $\approx 4592$\,deg$^2$) providing more accurate sky localizations for radio follow up, current-generation GW detectors and radio arrays would not be able to probe the mini-disk scenario. The accurate localizations provided by instruments like \textit{Swift}/BAT are crucial for enabling pointed VLA follow-up of the 10 BBHs per year that the O5 network is expected to detect within 630\,Mpc, the distance out to which the VLA could unveil BBH radio afterglows in the mini-disk scenario (dashed horizontal line in Figure\,\ref{fig:sims}).
The ngVLA in survey mode, by contrast, could follow up BBH events without requiring more accurate localizations from independent EM facilities. In the O5 BBH population, 27 events per year fall within the ngVLA survey capabilities, and the next-generation CE-ET network improves on this substantially, with 83 events per year falling within the ngVLA survey-mode region.
Remarkably, the next-generation GW networks localize events so precisely that the number of BBHs per year accessible to the ngVLA in \textit{pointed} mode exceeds that accessible in survey mode: CE-ET could localize 245 (8) BBHs per year at luminosity distances that would allow the ngVLA (VLA) to probe the mini-disk scenario, to sky areas well matched to the FOV of the ngVLA (VLA). This result is even more exciting when considering that ngVLA's coverage of the northern hemisphere will be complemented by facilities in the southern hemisphere like the Square Kilometer Array (SKA), with a 1\,deg field of view and nominal RMS of 2\,$\mu$Jy at 1.4 GHz \citep{SKAPerformance}.

\section{Summary}
\label{sec:summary}

We presented VLA 3\,GHz follow-up of the 85\% confidence region of a sub-threshold BAT-GUANO GRB that was temporally and spatially coincident with the gravitational-wave source GW241125, together with an analysis of \textit{Swift}/XRT observations and available optical constraints. Our radio observations identified 25 sources in total, 5 of which are variable, but none exhibit the evolution expected for a GRB afterglow from a BBH mini-disk jet. We also report a radio counterpart to the only confident X-ray source within the BAT-GUANO region, also consistent with a EP candidate counterpart, which in the radio is consistent with persistent emission. 

In the absence of a transient radio counterpart identification, and in the context of the mini-disk scenario for BBHs, our radio observations constrain the isotropic-equivalent kinetic energy of any relativistic ejecta to $\lesssim 3\times10^{50}$\,erg for a fiducial density $n_{\rm ISM} = 1$\,cm$^{-3}$. 

Finally, we have shown how probing the mini-disk scenario for stellar-mass BBHs depends critically on how well events can be localized. For current-generation GW detectors, this requires external localization from large-field high-energy instruments such as \textit{Swift}/BAT: only with such constraints can pointed VLA observations follow up the $\sim10$ BBHs per year that the O5 network is expected to detect within 630\,Mpc, the VLA's reach for BBH radio afterglows in this scenario. The ngVLA in survey mode relaxes this requirement, capturing 27 such events per year from the O5 population without independent EM localization, rising to 83 per year for the next-generation CE-ET network. Most strikingly, the localization precision of next-generation GW networks is good enough that pointed ngVLA follow-up outperforms survey mode: the CE-ET network could localize 245 (8) BBHs per year at distances accessible to the ngVLA (VLA) and to sky areas matched to these facilities' small FOV. With northern-hemisphere ngVLA coverage complemented by southern facilities such as the SKA, the prospects for systematically probing relativistic ejecta from BBH mergers are rather promising.

\begin{acknowledgments}
N.G. and A.C. acknowledge support from the National Science Foundation (NSF) via grant AST-2431072. S.R. acknowledges support from the Astrophysics Center for Multi-messenger Studies in Europe (ACME), funded under the European Union’s Horizon Europe Research and Innovation Program, Grant Agreement No. 101131928. DJ acknowledges the Science and Technology Facilities Council (STFC) for support through grants ST/V005618/1 and ST/Y004272/1. The National Radio Astronomy Observatory and Green Bank Observatory are facilities of the U.S. National Science Foundation operated under cooperative agreement by Associated Universities, Inc. Additionally, we thank Douglas Cowen, Aaron Tohuvavohu, Chad Hanna, Bangalore Sathyaprakash, Avery Eddins, Hugo Ayala, Stephanie Wissel, Felicia McBride, and Jamie Kennea for their expertise and contributions to the observation proposal and Avery Eddins for her feedback on the text and figures of this paper. 

\end{acknowledgments}

\vspace{5mm}

\appendix
\section{Comparing the VLA Automated Imaging Pipeline with Manual Imaging in \texttt{CASA}}
The VLA images were reduced using the recently implemented automated imaging pipeline in \texttt{CASA} \citep{VLAPipeline, ALMAPipeline}. To quantify the differences between this pipeline and a manual reduction, we imaged the third VLA epoch (E3) by hand and compared the two. Using the mask files from each imaging (automated and by hand) procedure, we compiled a list of sources identified in both reductions and analyzed those falling within our the \textit{Swift}/GUANO localization region of the candidate $\gamma$-ray counterpart to GW241125. Above SNR\,=7 the two lists agree completely, with increasing discrepancies toward lower SNR: at SNR\,=6 only one source recovered manually was missed by the pipeline, while at our base cutoff of SNR\,=5
the pipeline missed 11 sources. The non-primary-beam-corrected RMS noise is consistent between the two images ($\approx 3.8\,\mu$Jy for the pipeline image versus $\approx 3.7\,\mu$Jy for the manual one), and the peak fluxes of individual sources agree within their uncertainties.
For our purposes the automated imaging pipeline is sufficient: the confidence in any individual SNR$\approx 5$ source is in any case weakened by the large localization region we are searching as related to the GW localization uncertainties. Although the pipeline catalog is incomplete below SNR\,$\approx7$, it recovers at least 75\% of sources automatically and provides a reliable basis for targeted manual reduction to recover the remainder, making the overall process substantially more efficient. 

\section{Full Source List} 
In Table 5, we report the full list of radio sources found in the \textit{Swift}/BAT region that reach an SNR of 7 in any of the first three epochs.

\begin{deluxetable*}{ccccccc}[]

\tablecolumns{7}
\tablewidth{1.0\columnwidth} 
\tablecaption{Radio sources identified in the VLA images of the field of the \textit{Swift}/GUANO candidate counterpart to GW241125.} 
\tablehead{ \colhead{\textbf{Source}} & \colhead{\textbf{Epoch}} & \colhead{\textbf{RMS}} & \colhead{\textbf{R.A. Dec.}} & \colhead{\textbf{Pos. Err}} & \colhead{\textbf{Peak F$_\nu$}} & \colhead{\textbf{E1 Offset}} \\ \colhead{} & \colhead{(days)} & \colhead{($\mu$Jy)} & \colhead{(hh:mm:ss deg:mm:ss)} & \colhead{($\arcsec$)} & \colhead{($\mu$Jy)} & \colhead{($\arcsec$)}}
\startdata
    S1       & 5  & 5.45 & 03:51:53.771 +69:44:48.00 & 0.15 & 45.5 $\pm$ 6.0 & 0  \\
    $\cdots$ & 18 & 5.44 & $\cdots$                  & 0.15 & 45.1 $\pm$ 5.9 & 0.14 \\
    $\cdots$  & 46 & 5.64 & $\cdots$                  & 0.14 & 51.0 $\pm$ 6.2 & 0.014 \\
    \hline
    S2        & 5  & 5.35 & 03:51:58.166 +69:44:51.64 & 0.11 & 106.3 $\pm$ 7.5 & 0  \\
    $\cdots$  & 18 & 5.35 & $\cdots$                  & 0.12 & 75.3  $\pm$ 6.6 & 0.10 \\
    $\cdots$  & 46 & 5.54 & $\cdots$                  & 0.11 & 112.1 $\pm$ 7.9 & 0.11 \\
    \hline
    S3        & 5  & 5.27 & 03:52:52.685 +69:43:47.75 & 0.11 & 92.3 $\pm$ 7.0 & 0  \\
    $\cdots$  & 18 & 5.26 & $\cdots$                  & 0.12 & 85.9 $\pm$ 6.8 & 0.14 \\
    $\cdots$  & 46 & 5.46 & $\cdots$                  & 0.11 & 91.0 $\pm$ 7.1 & 0.079 \\
    \hline
    S4        & 5  & 4.38 & 03:52:14.269 +69:43:00.31 & 0.16 & 34.0 $\pm$ 4.7 & 0  \\
    $\cdots$  & 18 & 4.37 & $\cdots$                  & 0.17 & 30.0 $\pm$ 4.6 & 0.10 \\
    $\cdots$  & 46 & 4.53 & $\cdots$                  & 0.16 & 30.9 $\pm$ 4.8 & 0.023 \\
    \hline
    S5        & 5  & 4.56 & 03:52:01.420 +69:43:07.56 & 0.16 & 35.3 $\pm$ 4.9 & 0  \\
    $\cdots$  & 18 & 4.55 & $\cdots$                  & 0.17 & 31.2 $\pm$ 4.8 & 0.20 \\
    $\cdots$  & 46 & 4.72 & $\cdots$                  & 0.16 & 33.2 $\pm$ 5.0 & 0.10 \\
    \hline
    S6        & 5  & 4.60 & 03:51:57.104 +69:42:54.66 & 0.11 & 131.3 $\pm$ 8.1 & 0  \\
    $\cdots$  & 18 & 4.59 & $\cdots$                  & 0.11 & 119.3 $\pm$ 7.7 & 0.10 \\
    $\cdots$  & 46 & 4.76 & $\cdots$                  & 0.11 & 109.4 $\pm$ 7.2 & 0.068 \\
    \hline
    S7        & 5  & 4.88 & 03:51:46.507 +69:42:40.73 & 0.16 & 37.5 $\pm$ 5.2 & 0 \\
    $\cdots$  & 18 & 4.87 & $\cdots$                  & 0.15 & 40.1 $\pm$ 5.3 & 0.20 \\
    $\cdots$  & 46 & 5.05 & $\cdots$                  & 0.16 & 34.1 $\pm$ 5.3 & 0.17 \\
    \hline
    S8        & 5  & 4.90 & 03:51:46.198 +69:42:44.43 & 0.14 & 43.2 $\pm$ 5.3 & 0  \\
    $\cdots$  & 18 & 4.89 & $\cdots$                  & 0.13 & 54.9 $\pm$ 5.5 & 0  \\
    $\cdots$  & 46 & 5.08 & $\cdots$                  & 0.13 & 57.2 $\pm$ 5.7 & 0.13 \\
    \hline
    S9        & 5  & 5.84 & 03:51:32.015 +69:43:23.63 & 0.10 & 200 $\pm$ 11 & 0  \\
    $\cdots$  & 18 & 5.84 & $\cdots$                  & 0.10 & 199 $\pm$ 12 & 0.14 \\
    $\cdots$  & 46 & 6.05 & $\cdots$                  & 0.10 & 192 $\pm$ 11 & 0.044 \\
    \hline
    S10       & 5  & 4.64 & 03:51:50.370 +69:41:52.97 & 0.17 & 32.3 $\pm$ 4.9 & 0  \\
    $\cdots$  & 18 & 4.62 & $\cdots$                  & 0.15 & 39.3 $\pm$ 5.0 & 0.10 \\
    $\cdots$  & 46 & 4.80 & $\cdots$                  & 0.15 & 37.2 $\pm$ 5.2 & 0.026 \\
    \hline
    S11       & 5  & 4.37 & 03:52:04.314 +69:42:20.98 & 0.17 & 30.2 $\pm$ 4.6 & 0  \\
    $\cdots$  & 18 & 4.35 & $\cdots$                  & 0.27 & 16.9 $\pm$ 4.4 & 0.14 \\
    $\cdots$  & 46 & 4.52 & $\cdots$                  & 0.15 & 33.3 $\pm$ 4.8 & 0.15 \\
    \hline
    S12       & 5  & 4.33 & 03:52:05.926 +69:42:12.45 & 0.12 & 65.6 $\pm$ 5.4 & 0  \\
    $\cdots$  & 18 & 4.31 & $\cdots$                  & 0.12 & 69.6 $\pm$ 5.5 & 0.010 \\
    $\cdots$  & 46 & 4.48 & $\cdots$                  & 0.11 & 74.3 $\pm$ 5.9 & 0.090\\
    \hline
    S13       & 5  & 4.20 & 03:52:19.710 +69:41:17.52 & 0.13 & 42.8 $\pm$ 4.7 & 0  \\
    $\cdots$  & 18 & 4.18 & $\cdots$                  & 0.16 & 31.7 $\pm$ 4.5 & 0.10 \\
    $\cdots$  & 46 & 4.34 & $\cdots$                  & 0.15 & 34.0 $\pm$ 4.7 & 0.070 \\
    \hline
    S14       & 5  & 5.35 & 03:52:56.091 +69:39:11.18 & 0.11 & 99.9 $\pm$ 7.3 & 0  \\
    $\cdots$  & 18 & 5.34 & $\cdots$                  & 0.11 & 96.7 $\pm$ 7.2 & 0.14 \\
    $\cdots$  & 46 & 5.54 & $\cdots$                  & 0.11 & 93.7 $\pm$ 7.1 & 0.010 \\
    \hline
    S15       & 5  & 4.83 & 03:52:52.560 +69:40:41.22 & 0.11 & 129.8 $\pm$ 7.6 & 0  \\
    $\cdots$  & 18 & 4.82 & $\cdots$                  & 0.11 & 142.0 $\pm$ 8.1 & 0 \\
    $\cdots$  & 46 & 5.00 & $\cdots$                  & 0.11 & 125.7 $\pm$ 7.6 & 0.023 \\
    \hline
\enddata  
\end{deluxetable*}

\begin{deluxetable*}{cccccccc}[]

\tablecolumns{8}
\tablewidth{1.0\columnwidth} 
\startdata
    S16       & 5  & 4.36 & 03:52:18.251 +69:39:49.52 & 0.16 & 32.1 $\pm$ 4.6 & 0  \\
    $\cdots$  & 18 & 4.35 & $\cdots$                  & 0.16 & 32.4 $\pm$ 4.6 & 0.14 \\
    $\cdots$  & 46 & 4.51 & $\cdots$                  & 0.14 & 40.2 $\pm$ 5.0 & 0.16 \\
    \hline
    S17       & 5  & 4.39 & 03:52:23.142 +69:39:42.81 & 0.13 & 50.5 $\pm$ 5.0 & 0  \\
    $\cdots$  & 18 & 4.38 & $\cdots$                  & 0.13 & 52.5 $\pm$ 5.1 & 0  \\
    $\cdots$  & 46 & 4.55 & $\cdots$                  & 0.17 & 45.1 $\pm$ 5.1 & 0.13 \\
    \hline
    S18       & 5  & 4.60 & 03:52:23.754 +69:38:59.91 & 0.15 & 38.1 $\pm$ 5.0 & 0  \\
    $\cdots$  & 18 & 4.59 & $\cdots$                  & 0.20 & 25.1 $\pm$ 4.8 & 0.10 \\
    $\cdots$  & 46 & 4.76 & $\cdots$                  & 0.17 & 31.0 $\pm$ 5.0 & 0.068 \\
    \hline
    S19       & 5  & 5.11 & 03:52:24.508 +69:37:54.11 & 0.11 & 109.6 $\pm$ 7.5 & 0  \\
    $\cdots$  & 18 & 5.10 & $\cdots$                  & 0.11 & 104.3 $\pm$ 7.3 & 0.15 \\ 
    $\cdots$  & 46 & 5.29 & $\cdots$                  & 0.11 & 116.8 $\pm$ 7.9 & 0.079 \\
    \hline
    S20       & 5  & 5.50 & 03:52:37.229 +69:37:35.16 & 0.16 & 40.9 $\pm$ 5.9 & 0  \\
    $\cdots$  & 18 & 5.50 & $\cdots$                  & 0.15 & 49.0 $\pm$ 6.0 & 0.22 \\
    $\cdots$  & 46 & 5.70 & $\cdots$                  & 0.15 & 43.1 $\pm$ 6.1 & 0.056 \\
    \hline
    S21       & 5  & 5.44 & 03:52:21.726 +69:37:22.09 & 0.12 & 87.4  $\pm$ 7.1 & 0  \\
    $\cdots$  & 18 & 5.43 & $\cdots$                  & 0.12 & 73.4  $\pm$ 6.6 & 0.099 \\
    $\cdots$  & 46 & 5.63 & $\cdots$                  & 0.11 & 108.7 $\pm$ 7.9 & 0.13 \\
    \hline
    S22       & 5  & 5.36 & 03:52:56.091 +69:39:09.55 & 0.13 & 65.6 $\pm$ 6.2 & 0  \\
    $\cdots$  & 18 & 5.35 & $\cdots$                  & 0.13 & 67.0 $\pm$ 6.3 & 0 \\
    $\cdots$  & 46 & 5.55 & $\cdots$                  & 0.12 & 64.3 $\pm$ 6.4 & 0.051 \\
    \hline
    S23       & 5  & 4.63 & 03:52:23.753 +69:43:51.21 & 0.22 & 21.5 $\pm$ 4.8 & 0  \\
    $\cdots$  & 18 & 4.61 & $\cdots$                  & 0.33 & 14.3 $\pm$ 4.6 & 0.20 \\
    $\cdots$  & 46 & 4.79 & $\cdots$                  & 0.15 & 34.9 $\pm$ 5.1 & 0.16 \\
    \hline
    S24       & 5  & 5.34 & 03:53:03.815 +69:41:33.96 & 0.16 & 38.9 $\pm$ 5.7 & 0  \\
    $\cdots$  & 18 & 5.34 & $\cdots$                  & 0.26 & 21.2 $\pm$ 5.5 & 0.32 \\
    $\cdots$  & 46 & 5.53 & $\cdots$                  & 0.23 & 23.3 $\pm$ 5.6 & 0.12 \\
    \hline
    S25       & 5  & 5.20 & 03:52:54.129 +69:39:20.50 & 0.13 & 55.1 $\pm$ 5.9 & 0  \\
    $\cdots$  & 18 & 5.19 & $\cdots$                  & 0.15 & 46.8 $\pm$ 5.7 & 0  \\
    $\cdots$  & 46 & 5.39 & $\cdots$                  & 0.20 & 25.9 $\pm$ 5.5 & 0.26 \\
\enddata  
\tablecomments{Columns in this table are: source name, epoch, time since BBH merger, RMS at source location, RA Dec, position error, peak fluxes, offset from epoch 1 position. The rows correspond to the source's evolution through the first three epochs (E1-E3), during which the VLA was in its  A configuration (highest resolution images).}
\label{tab:sourcesall2}
\end{deluxetable*}


\bibliography{S241125n}{}

@misc{pesummary,
 author = "Hoy, Charlie",
 howpublished = {\url{https://lscsoft.docs.ligo.org/pesummary/stable/gw/parameters.html}},
 title = {PESummary},
 year = {2021}
}

@article{Divyajyoti:2026gds,
    author = {{Divyajyoti} and {Fairhurst}, Stephen and {Hannam}, Mark and {Singh}, Mukesh Kumar},
        title = "{Mapping the star formation peak with LIGO A\# and Next-Generation detectors}",
      journal = {arXiv e-prints},
         year = 2026,
        month = jun,
          eid = {arXiv:2606.05151},
        pages = {arXiv:2606.05151},
          doi = {10.48550/arXiv.2606.05151},
archivePrefix = {arXiv},
       eprint = {2606.05151},
 primaryClass = {gr-qc},
       adsurl = {https://ui.adsabs.harvard.edu/abs/2026arXiv260605151D}
}

@misc{VLAPipeline,
  author = {NRAO},
  title = {{VLA Imaging Pipeline}},
  howpublished = "\url{https://science.nrao.edu/facilities/vla/data-processing/pipeline/vipl_666}",
  year = {2025}, 
  note = "[Online]"
}

@INPROCEEDINGS{2018IAUS..336..426M,
       author = {{Murphy}, Eric J.},
        title = "{A next-generation Very Large Array}",
    booktitle = {Astrophysical Masers: Unlocking the Mysteries of the Universe},
         year = 2018,
       editor = {{Tarchi}, A. and {Reid}, M.~J. and {Castangia}, P.},
       series = {IAU Symposium},
       volume = {336},
        month = aug,
        pages = {426-432},
          doi = {10.1017/S1743921317009838},
archivePrefix = {arXiv},
       eprint = {1711.09921},
 primaryClass = {astro-ph.IM},
       adsurl = {https://ui.adsabs.harvard.edu/abs/2018IAUS..336..426M}
}

@ARTICLE{ALMAPipeline,
       author = {{Hunter}, Todd R. and {Indebetouw}, Remy and {Brogan}, Crystal L. and {Berry}, Kristin and {Chang}, Chin-Shin and {Francke}, Harold and {Geers}, Vincent C. and {G{\'o}mez}, Laura and {Hibbard}, John E. and {Humphreys}, Elizabeth M. and {Kent}, Brian R. and {Kepley}, Amanda A. and {Kunneriath}, Devaky and {Lipnicky}, Andrew and {Loomis}, Ryan A. and {Mason}, Brian S. and {Masters}, Joseph S. and {Maud}, Luke T. and {Muders}, Dirk and {Sabater}, Jose and {Sugimoto}, Kanako and {Sz{\H{u}}cs}, L{\'a}szl{\'o} and {Vasiliev}, Eugene and {Videla}, Liza and {Villard}, Eric and {Williams}, Stewart J. and {Xue}, Rui and {Yoon}, Ilsang},
        title = "{The ALMA Interferometric Pipeline Heuristics}",
      journal = {\pasp},
         year = 2023,
        month = jul,
       volume = {135},
       number = {1049},
          eid = {074501},
        pages = {074501},
          doi = {10.1088/1538-3873/ace216},
archivePrefix = {arXiv},
       eprint = {2306.07420},
 primaryClass = {astro-ph.IM},
       adsurl = {https://ui.adsabs.harvard.edu/abs/2023PASP..135g4501H}
}

@ARTICLE{SKAPerformance,
       author = {{Braun}, Robert and {Bonaldi}, Anna and {Bourke}, Tyler and {Keane}, Evan and {Wagg}, Jeff},
        title = "{Anticipated Performance of the Square Kilometre Array -- Phase 1 (SKA1)}",
      journal = {arXiv e-prints},
         year = 2019,
        month = dec,
          eid = {arXiv:1912.12699},
        pages = {arXiv:1912.12699},
          doi = {10.48550/arXiv.1912.12699},
archivePrefix = {arXiv},
       eprint = {1912.12699},
 primaryClass = {astro-ph.IM},
       adsurl = {https://ui.adsabs.harvard.edu/abs/2019arXiv191212699B}
}

@ARTICLE{KaraP,
       author = {{Merfeld}, Kara and {Corsi}, Alessandra},
        title = "{Probing Binary Neutron Star Merger Ejecta and Remnants with Gravitational Wave and Radio Observations}",
      journal = {arXiv e-prints},
         year = 2025,
        month = jun,
          eid = {arXiv:2506.22835},
        pages = {arXiv:2506.22835},
          doi = {10.48550/arXiv.2506.22835},
archivePrefix = {arXiv},
       eprint = {2506.22835},
 primaryClass = {astro-ph.HE},
       adsurl = {https://ui.adsabs.harvard.edu/abs/2025arXiv250622835M}
}

@misc{O5c_sensitivity,
    author = {{{LIGO Scientific Collaboration}}},
    title = {A+/O5 strain curve projections},
    url = {https://dcc.ligo.org/LIGO-T2500310/public},
    year = {2025}
}

@misc{CE_sensitivity,
    author = {{{CE Consortium}}},
    title = {Cosmic Explorer Strain Sensitivity},
    url = {https://dcc.cosmicexplorer.org/CE-T2000017-v8/public},
    year = {2024}
}

@misc{ET_sensitivity_CoBA,
    author = {{{ET Collaboration}}},
    title = {ET sensitivity curves used for CoBA Science Study},
    url = {https://apps.et-gw.eu/tds/?r=18213},
    year = {2023}
}

@article{VIRGO:2014yos,
 archiveprefix = {arXiv},
 author = {Acernese, F. and others},
 collaboration = {VIRGO},
 doi = {10.1088/0264-9381/32/2/024001},
 eprint = {1408.3978},
 journal = {Class. Quant. Grav.},
 number = {2},
 pages = {024001},
 primaryclass = {gr-qc},
 title = {{Advanced Virgo: a second-generation interferometric gravitational wave detector}},
 volume = {32},
 year = {2015}
}

@article{ET:2025xjr,
    author = "Abac, Adrian and others",
    title = "{The Science of the Einstein Telescope}",
      journal = {\jcap},
         year = 2026,
        month = mar,
       volume = {2026},
       number = {3},
          eid = {081},
        pages = {081},
          doi = {10.1088/1475-7516/2026/03/081},
archivePrefix = {arXiv},
       eprint = {2503.12263},
 primaryClass = {gr-qc},
       adsurl = {https://ui.adsabs.harvard.edu/abs/2026JCAP...03..081A}
}

@article{Ofek_2011,
doi = {10.1088/0004-637X/740/2/65},
url = {https://doi.org/10.1088/0004-637X/740/2/65},
year = {2011},
month = {sep},
publisher = {The American Astronomical Society},
volume = {740},
number = {2},
pages = {65},
author = {Ofek, E. O. and Frail, D. A. and Breslauer, B. and Kulkarni, S. R. and Chandra, P. and Gal-Yam, A. and Kasliwal, M. M. and Gehrels, N.},
title = {A VERY LARGE ARRAY SEARCH FOR 5 GHz RADIO TRANSIENTS AND VARIABLES AT LOW GALACTIC LATITUDES},
journal = {The Astrophysical Journal}
}

@article{Stern_2012,
doi = {10.1088/0004-637X/753/1/30},
url = {https://doi.org/10.1088/0004-637X/753/1/30},
year = {2012},
month = {jun},
publisher = {The American Astronomical Society},
volume = {753},
number = {1},
pages = {30},
author = {Stern, Daniel and Assef, Roberto J. and Benford, Dominic J. and Blain, Andrew and Cutri, Roc and Dey, Arjun and Eisenhardt, Peter and Griffith, Roger L. and Jarrett, T. H. and Lake, Sean and Masci, Frank and Petty, Sara and Stanford, S. A. and Tsai, Chao-Wei and Wright, E. L. and Yan, Lin and Harrison, Fiona and Madsen, Kristin},
title = {MID-INFRARED SELECTION OF ACTIVE GALACTIC NUCLEI WITH THE WIDE-FIELD INFRARED SURVEY EXPLORER. I. CHARACTERIZING WISE-SELECTED ACTIVE GALACTIC NUCLEI IN COSMOS},
journal = {The Astrophysical Journal}
}

@article{Tabatabaei_2017,
doi = {10.3847/1538-4357/836/2/185},
url = {https://doi.org/10.3847/1538-4357/836/2/185},
year = {2017},
month = {feb},
publisher = {The American Astronomical Society},
volume = {836},
number = {2},
pages = {185},
author = {Tabatabaei, F. S. and Schinnerer, E. and Krause, M. and Dumas, G. and Meidt, S. and Damas-Segovia, A. and Beck, R. and Murphy, E. J. and Mulcahy, D. D. and Groves, B. and Bolatto, A. and Dale, D. and Galametz, M. and Sandstrom, K. and Boquien, M. and Calzetti, D. and Kennicutt, R. C. and Hunt, L. K. and Looze, I. De and Pellegrini, E. W.},
title = {The Radio Spectral Energy Distribution and Star-formation Rate Calibration in Galaxies},
journal = {The Astrophysical Journal}
}

@ARTICLE{1994ApJS...95....1E,
       author = {{Elvis}, Martin and {Wilkes}, Belinda J. and {McDowell}, Jonathan C. and {Green}, Richard F. and {Bechtold}, Jill and {Willner}, S.~P. and {Oey}, M.~S. and {Polomski}, Elisha and {Cutri}, Roc},
        title = "{Atlas of Quasar Energy Distributions}",
      journal = {\apjs},
         year = 1994,
        month = nov,
       volume = {95},
        pages = {1},
          doi = {10.1086/192093},
       adsurl = {https://ui.adsabs.harvard.edu/abs/1994ApJS...95....1E}
}

@ARTICLE{2006A&A...451...35B,
       author = {{Balmaverde}, B. and {Capetti}, A. and {Grandi}, P.},
        title = "{The Chandra view of the 3C/FR I sample of low luminosity radio-galaxies}",
      journal = {\aap},
         year = 2006,
        month = may,
       volume = {451},
       number = {1},
        pages = {35-44},
          doi = {10.1051/0004-6361:20053799},
archivePrefix = {arXiv},
       eprint = {astro-ph/0601175},
 primaryClass = {astro-ph},
       adsurl = {https://ui.adsabs.harvard.edu/abs/2006A&A...451...35B}
}

@ARTICLE{GRRev,
       author = {{Kumar}, Pawan and {Zhang}, Bing},
        title = "{The physics of gamma-ray bursts \& relativistic jets}",
      journal = {\physrep},
         year = 2015,
        month = feb,
       volume = {561},
        pages = {1-109},
          doi = {10.1016/j.physrep.2014.09.008},
archivePrefix = {arXiv},
       eprint = {1410.0679},
 primaryClass = {astro-ph.HE},
       adsurl = {https://ui.adsabs.harvard.edu/abs/2015PhR...561....1K}
}

@ARTICLE{PiranRev,
       author = {{Piran}, Tsvi},
        title = "{The physics of gamma-ray bursts}",
      journal = {Reviews of Modern Physics},
         year = 2004,
        month = oct,
       volume = {76},
       number = {4},
        pages = {1143-1210},
          doi = {10.1103/RevModPhys.76.1143},
archivePrefix = {arXiv},
       eprint = {astro-ph/0405503},
 primaryClass = {astro-ph},
       adsurl = {https://ui.adsabs.harvard.edu/abs/2004RvMP...76.1143P}
}

@ARTICLE{Sari98,
       author = {{Sari}, Re'em and {Piran}, Tsvi and {Narayan}, Ramesh},
        title = "{Spectra and Light Curves of Gamma-Ray Burst Afterglows}",
      journal = {\apjl},
         year = 1998,
        month = apr,
       volume = {497},
       number = {1},
        pages = {L17-L20},
          doi = {10.1086/311269},
archivePrefix = {arXiv},
       eprint = {astro-ph/9712005},
 primaryClass = {astro-ph},
       adsurl = {https://ui.adsabs.harvard.edu/abs/1998ApJ...497L..17S}
}

@ARTICLE{FongGRBs,
       author = {{Fong}, W. and {Berger}, E. and {Margutti}, R. and {Zauderer}, B.~A.},
        title = "{A Decade of Short-duration Gamma-Ray Burst Broadband Afterglows: Energetics, Circumburst Densities, and Jet Opening Angles}",
      journal = {\apj},
         year = 2015,
        month = dec,
       volume = {815},
       number = {2},
          eid = {102},
        pages = {102},
          doi = {10.1088/0004-637X/815/2/102},
archivePrefix = {arXiv},
       eprint = {1509.02922},
 primaryClass = {astro-ph.HE},
       adsurl = {https://ui.adsabs.harvard.edu/abs/2015ApJ...815..102F}
}

@ARTICLE{GhirlandaGRBs,
       author = {{Ghirlanda}, Giancarlo and {Ghisellini}, Gabriele and {Lazzati}, Davide},
        title = "{The Collimation-corrected Gamma-Ray Burst Energies Correlate with the Peak Energy of Their {\ensuremath{\nu}}F$_{{\ensuremath{\nu}}}$ Spectrum}",
      journal = {\apj},
         year = 2004,
        month = nov,
       volume = {616},
       number = {1},
        pages = {331-338},
          doi = {10.1086/424913},
archivePrefix = {arXiv},
       eprint = {astro-ph/0405602},
 primaryClass = {astro-ph},
       adsurl = {https://ui.adsabs.harvard.edu/abs/2004ApJ...616..331G}
}

@article{Jarrett_2017,
doi = {10.3847/1538-4357/836/2/182},
url = {https://doi.org/10.3847/1538-4357/836/2/182},
year = {2017},
month = {feb},
publisher = {The American Astronomical Society},
volume = {836},
number = {2},
pages = {182},
author = {Jarrett, T. H. and Cluver, M. E. and Magoulas, C. and Bilicki, M. and Alpaslan, M. and Bland-Hawthorn, J. and Brough, S. and Brown, M. J. I. and Croom, S. and Driver, S. and Holwerda, B. W. and Hopkins, A. M. and Loveday, J. and Norberg, P. and Peacock, J. A. and Popescu, C. C. and Sadler, E. M. and Taylor, E. N. and Tuffs, R. J. and Wang, L.},
title = {Galaxy and Mass Assembly (GAMA): Exploring the WISE Web in G12},
journal = {The Astrophysical Journal}
}

@ARTICLE{sGRBs,
       author = {{Berger}, Edo},
        title = "{Short-Duration Gamma-Ray Bursts}",
      journal = {\araa},
         year = 2014,
        month = aug,
       volume = {52},
        pages = {43-105},
          doi = {10.1146/annurev-astro-081913-035926},
archivePrefix = {arXiv},
       eprint = {1311.2603},
 primaryClass = {astro-ph.HE},
       adsurl = {https://ui.adsabs.harvard.edu/abs/2014ARA&A..52...43B}
}

@ARTICLE{2020FrASS...7...78L,
       author = {{Lazzati}, Davide},
        title = "{Short Duration Gamma-Ray Bursts and Their Outflows in Light of GW170817}",
      journal = {Frontiers in Astronomy and Space Sciences},
         year = 2020,
        month = nov,
       volume = {7},
          eid = {78},
        pages = {78},
          doi = {10.3389/fspas.2020.578849},
archivePrefix = {arXiv},
       eprint = {2009.01773},
 primaryClass = {astro-ph.HE},
       adsurl = {https://ui.adsabs.harvard.edu/abs/2020FrASS...7...78L}
}

@ARTICLE{RegaladeCat,
       author = {{Tranin}, Hugo and {Blagorodnova}, Nadejda and {G{\'o}mez-Mu{\~n}oz}, Marco A. and {Wavasseur}, Maxime and {Groot}, Paul J. and {Landsberg}, Lloyd and {Stoppa}, Fiorenzo and {Bloemen}, Steven and {Vreeswijk}, Paul M. and {Pieterse}, Dani{\"e}lle L.~A. and {van Roestel}, Jan and {Scaringi}, Simone and {Faris}, Sara},
        title = "{A catalog to unite them all: REGALADE, a revised galaxy compilation for the advanced detector era}",
      journal = {\aap},
         year = 2026,
        month = feb,
       volume = {706},
          eid = {A284},
        pages = {A284},
          doi = {10.1051/0004-6361/202556896},
archivePrefix = {arXiv},
       eprint = {2508.13267},
 primaryClass = {astro-ph.GA},
       adsurl = {https://ui.adsabs.harvard.edu/abs/2026A&A...706A.284T}
}

@ARTICLE{PernaHorizons,
       author = {{Perna}, Rosalba and {Chruslinska}, Martyna and {Corsi}, Alessandra and {Belczynski}, Krzysztof},
        title = "{Binary black hole mergers within the LIGO horizon: statistical properties and prospects for detecting electromagnetic counterparts}",
      journal = {\mnras},
         year = 2018,
        month = jul,
       volume = {477},
       number = {3},
        pages = {4228-4240},
          doi = {10.1093/mnras/sty814},
archivePrefix = {arXiv},
       eprint = {1708.09402},
 primaryClass = {astro-ph.HE},
       adsurl = {https://ui.adsabs.harvard.edu/abs/2018MNRAS.477.4228P}
}

@INPROCEEDINGS{2020ASPC..527..571K,
       author = {{Kent}, B.~R. and {Masters}, J.~S. and {Chandler}, C.~J. and {Tobin}, J.~J. and {Marvil}, J. and {Ott}, J. and {Myers}, S.~T. and {Medlin}, D. and {Kimball}, A.~E. and {Schinzel}, F.~K. and {Lacy}, M. and {Kern}, J. and {Butler}, B.~J. and {Sugimoto}, K. and {Muders}, D. and {Williams}, S.~J. and {Geers}, V.~G.},
        title = "{Pipeline Calibration and Imaging for the Very Large Array}",
    booktitle = {Astronomical Data Analysis Software and Systems XXIX},
         year = 2020,
       editor = {{Pizzo}, R. and {Deul}, E.~R. and {Mol}, J.~D. and {de Plaa}, J. and {Verkouter}, H.},
       series = {Astronomical Society of the Pacific Conference Series},
       volume = {527},
        month = jan,
        pages = {571},
       adsurl = {https://ui.adsabs.harvard.edu/abs/2020ASPC..527..571K}
}

@ARTICLE{2026arXiv260527225T,
       author = {{The LIGO Scientific Collaboration} and {the Virgo Collaboration} and {the KAGRA Collaboration}},
        title = "{GWTC-5.0: Observations from the Second Part of the Fourth LIGO-Virgo-KAGRA Observing Run and Updates to the Gravitational-Wave Transient Catalog}",
      journal = {arXiv e-prints},
         year = 2026,
        month = may,
          eid = {arXiv:2605.27225},
        pages = {arXiv:2605.27225},
          doi = {10.48550/arXiv.2605.27225},
archivePrefix = {arXiv},
       eprint = {2605.27225},
 primaryClass = {gr-qc},
       adsurl = {https://ui.adsabs.harvard.edu/abs/2026arXiv260527225T}
}

@article{Chandra:2024dhf,
    author = "Chandra, Koustav",
    title = "{gwforge: a user-friendly package to generate gravitational-wave mock data}",
    eprint = "2407.21109",
    archivePrefix = "arXiv",
    primaryClass = "gr-qc",
    doi = "10.1088/1361-6382/ad9b68",
    journal = "Class. Quant. Grav.",
    volume = "42",
    number = "2",
    pages = "025003",
    year = "2025"
}

@article{KAGRA:2021duu,
 archiveprefix = {arXiv},
 author = {Abbott, R. and others},
 collaboration = {KAGRA, VIRGO, LIGO Scientific},
 doi = {10.1103/PhysRevX.13.011048},
 eprint = {2111.03634},
 journal = {Phys. Rev. X},
 number = {1},
 pages = {011048},
 primaryclass = {astro-ph.HE},
 reportnumber = {LIGO-P2100239 ; Data release: https://zenodo.org/record/5655785, LIGO-P2100239},
 title = {{Population of Merging Compact Binaries Inferred Using Gravitational Waves through GWTC-3}},
 volume = {13},
 year = {2023}
}

@article{Madau:2014bja,
    author = "Madau, Piero and Dickinson, Mark",
    title = "{Cosmic Star Formation History}",
    eprint = "1403.0007",
    archivePrefix = "arXiv",
    primaryClass = "astro-ph.CO",
    doi = "10.1146/annurev-astro-081811-125615",
    journal = "Ann. Rev. Astron. Astrophys.",
    volume = "52",
    pages = "415--486",
    year = "2014"
}

@article{Dominik:2013tma,
    author = "Dominik, Michal and Belczynski, Krzysztof and Fryer, Christopher and Holz, Daniel E. and Berti, Emanuele and Bulik, Tomasz and Mandel, Ilya and O'Shaughnessy, Richard",
    title = "{Double Compact Objects II: Cosmological Merger Rates}",
    eprint = "1308.1546",
    archivePrefix = "arXiv",
    primaryClass = "astro-ph.HE",
    doi = "10.1088/0004-637X/779/1/72",
    journal = "Astrophys. J.",
    volume = "779",
    pages = "72",
    year = "2013"
}

@article{Dominik:2012kk,
    author = "Dominik, Michal and Belczynski, Krzysztof and Fryer, Christopher and Holz, Daniel and Berti, Emanuele and Bulik, Tomasz and Mandel, Ilya and O'Shaughnessy, Richard",
    title = "{Double Compact Objects I: The Significance of the Common Envelope on Merger Rates}",
    eprint = "1202.4901",
    archivePrefix = "arXiv",
    primaryClass = "astro-ph.HE",
    doi = "10.1088/0004-637X/759/1/52",
    journal = "Astrophys. J.",
    volume = "759",
    pages = "52",
    year = "2012"
}

@article{Fishbach:2018edt,
    author = "Fishbach, Maya and Holz, Daniel E. and Farr, Will M.",
    title = "{Does the Black Hole Merger Rate Evolve with Redshift?}",
    eprint = "1805.10270",
    archivePrefix = "arXiv",
    primaryClass = "astro-ph.HE",
    doi = "10.3847/2041-8213/aad800",
    journal = "Astrophys. J. Lett.",
    volume = "863",
    number = "2",
    pages = "L41",
    year = "2018"
}

@ARTICLE{LazzatiAGNRadio,
       author = {{Wang}, Yi-Han and {Lazzati}, Davide and {Perna}, Rosalba},
        title = "{The emergence of diffused gamma-ray burst afterglows from the discs of active galactic nuclei}",
      journal = {\mnras},
         year = 2022,
        month = nov,
       volume = {516},
       number = {4},
        pages = {5935-5944},
          doi = {10.1093/mnras/stac1968},
archivePrefix = {arXiv},
       eprint = {2207.05084},
 primaryClass = {astro-ph.HE},
       adsurl = {https://ui.adsabs.harvard.edu/abs/2022MNRAS.516.5935W}
}

@ARTICLE{VariabilityStat,
       author = {{Mooley}, K.~P. and {Hallinan}, G. and {Bourke}, S. and {Horesh}, A. and {Myers}, S.~T. and {Frail}, D.~A. and {Kulkarni}, S.~R. and {Levitan}, D.~B. and {Kasliwal}, M.~M. and {Cenko}, S.~B. and {Cao}, Y. and {Bellm}, E. and {Laher}, R.~R.},
        title = "{The Caltech-NRAO Stripe 82 Survey (CNSS). I. The Pilot Radio Transient Survey In 50 deg$^{2}$}",
      journal = {\apj},
         year = 2016,
        month = feb,
       volume = {818},
       number = {2},
          eid = {105},
        pages = {105},
          doi = {10.3847/0004-637X/818/2/105},
archivePrefix = {arXiv},
       eprint = {1601.01693},
 primaryClass = {astro-ph.HE},
       adsurl = {https://ui.adsabs.harvard.edu/abs/2016ApJ...818..105M}
}

@ARTICLE{NVSS,
       author = {{Mauch}, Tom and {Sadler}, Elaine M.},
        title = "{Radio sources in the 6dFGS: local luminosity functions at 1.4 GHz for star-forming galaxies and radio-loud AGN}",
      journal = {\mnras},
         year = 2007,
        month = mar,
       volume = {375},
       number = {3},
        pages = {931-950},
          doi = {10.1111/j.1365-2966.2006.11353.x},
archivePrefix = {arXiv},
       eprint = {astro-ph/0612018},
 primaryClass = {astro-ph},
       adsurl = {https://ui.adsabs.harvard.edu/abs/2007MNRAS.375..931M}
}

@ARTICLE{Burrows+2005,
       author = {{Burrows}, David N. and {Hill}, J.~E. and {Nousek}, J.~A. and {Kennea}, J.~A. and {Wells}, A. and {Osborne}, J.~P. and {Abbey}, A.~F. and {Beardmore}, A. and {Mukerjee}, K. and {Short}, A.~D.~T. and {Chincarini}, G. and {Campana}, S. and {Citterio}, O. and {Moretti}, A. and {Pagani}, C. and {Tagliaferri}, G. and {Giommi}, P. and {Capalbi}, M. and {Tamburelli}, F. and {Angelini}, L. and {Cusumano}, G. and {Br{\"a}uninger}, H.~W. and {Burkert}, W. and {Hartner}, G.~D.},
        title = "{The Swift X-Ray Telescope}",
      journal = {\ssr},
         year = 2005,
        month = oct,
       volume = {120},
       number = {3-4},
        pages = {165-195},
          doi = {10.1007/s11214-005-5097-2},
archivePrefix = {arXiv},
       eprint = {astro-ph/0508071},
 primaryClass = {astro-ph},
       adsurl = {https://ui.adsabs.harvard.edu/abs/2005SSRv..120..165B}
}

@ARTICLE{AfterglowPy,
       author = {{Ryan}, Geoffrey and {van Eerten}, Hendrik and {Piro}, Luigi and {Troja}, Eleonora},
        title = "{Gamma-Ray Burst Afterglows in the Multimessenger Era: Numerical Models and Closure Relations}",
      journal = {\apj},
         year = 2020,
        month = jun,
       volume = {896},
       number = {2},
          eid = {166},
        pages = {166},
          doi = {10.3847/1538-4357/ab93cf},
archivePrefix = {arXiv},
       eprint = {1909.11691},
 primaryClass = {astro-ph.HE},
       adsurl = {https://ui.adsabs.harvard.edu/abs/2020ApJ...896..166R}
}

@ARTICLE{AGNWiseWedge,
       author = {{Mateos}, S. and {Alonso-Herrero}, A. and {Carrera}, F.~J. and {Blain}, A. and {Watson}, M.~G. and {Barcons}, X. and {Braito}, V. and {Severgnini}, P. and {Donley}, J.~L. and {Stern}, D.},
        title = "{Using the Bright Ultrahard XMM-Newton survey to define an IR selection of luminous AGN based on WISE colours}",
      journal = {\mnras},
         year = 2012,
        month = nov,
       volume = {426},
       number = {4},
        pages = {3271-3281},
          doi = {10.1111/j.1365-2966.2012.21843.x},
archivePrefix = {arXiv},
       eprint = {1208.2530},
 primaryClass = {astro-ph.CO},
       adsurl = {https://ui.adsabs.harvard.edu/abs/2012MNRAS.426.3271M}
}

@ARTICLE{VLASSTransients,
       author = {{Gordon}, Yjan A. and {Ferguson}, Peter S. and {Martinez}, Michael N. and {Hooper}, Eric J.},
        title = "{A Census of Variable Radio Sources at $3\,$GHz}",
      journal = {arXiv e-prints},
         year = 2025,
        month = aug,
          eid = {arXiv:2508.00976},
        pages = {arXiv:2508.00976},
          doi = {10.48550/arXiv.2508.00976},
archivePrefix = {arXiv},
       eprint = {2508.00976},
 primaryClass = {astro-ph.GA},
       adsurl = {https://ui.adsabs.harvard.edu/abs/2025arXiv250800976G}
}

@MISC{WISE1,
       author = {{Cutri}, R.~M. and {Wright}, E.~L. and {Conrow}, T. and {Fowler}, J.~W. and {Eisenhardt}, P.~R.~M. and {Grillmair}, C. and {Kirkpatrick}, J.~D. and {Masci}, F. and {McCallon}, H.~L. and {Wheelock}, S.~L. and {Fajardo-Acosta}, S. and {Yan}, L. and {Benford}, D. and {Harbut}, M. and {Jarrett}, T. and {Lake}, S. and {Leisawitz}, D. and {Ressler}, M.~E. and {Stanford}, S.~A. and {Tsai}, C.~W. and {Liu}, F. and {Helou}, G. and {Mainzer}, A. and {Gettings}, D. and {Gonzalez}, A. and {Hoffman}, D. and {Marsh}, K.~A. and {Padgett}, D. and {Skrutskie}, M.~F. and {Beck}, R.~P. and {Papin}, M. and {Wittman}, M.},
        title = "{Explanatory Supplement to the AllWISE Data Release Products}",
 howpublished = {Explanatory Supplement to the AllWISE Data Release Products, by R. M. Cutri et al.},
         year = 2013,
        month = nov,
        pages = {1},
       adsurl = {https://ui.adsabs.harvard.edu/abs/2013wise.rept....1C}
}

@ARTICLE{ChandraDeep,
       author = {{Mooley}, K.~P. and {Frail}, D.~A. and {Ofek}, E.~O. and {Miller}, N.~A. and {Kulkarni}, S.~R. and {Horesh}, A.},
        title = "{Sensitive Search for Radio Variables and Transients in the Extended Chandra Deep Field South}",
      journal = {\apj},
         year = 2013,
        month = may,
       volume = {768},
       number = {2},
          eid = {165},
        pages = {165},
          doi = {10.1088/0004-637X/768/2/165},
archivePrefix = {arXiv},
       eprint = {1303.6282},
 primaryClass = {astro-ph.CO},
       adsurl = {https://ui.adsabs.harvard.edu/abs/2013ApJ...768..165M}
}

@ARTICLE{GW170817Afterglow,
       author = {{Troja}, E. and {van Eerten}, H. and {Zhang}, B. and {Ryan}, G. and {Piro}, L. and {Ricci}, R. and {O'Connor}, B. and {Wieringa}, M.~H. and {Cenko}, S.~B. and {Sakamoto}, T.},
        title = "{A thousand days after the merger: Continued X-ray emission from GW170817}",
      journal = {\mnras},
         year = 2020,
        month = nov,
       volume = {498},
       number = {4},
        pages = {5643-5651},
          doi = {10.1093/mnras/staa2626},
archivePrefix = {arXiv},
       eprint = {2006.01150},
 primaryClass = {astro-ph.HE},
       adsurl = {https://ui.adsabs.harvard.edu/abs/2020MNRAS.498.5643T}
}

@ARTICLE{190521_EM,
       author = {{Graham}, M.~J. and {Ford}, K.~E.~S. and {McKernan}, B. and {Ross}, N.~P. and {Stern}, D. and {Burdge}, K. and {Coughlin}, M. and {Djorgovski}, S.~G. and {Drake}, A.~J. and {Duev}, D. and {Kasliwal}, M. and {Mahabal}, A.~A. and {van Velzen}, S. and {Belecki}, J. and {Bellm}, E.~C. and {Burruss}, R. and {Cenko}, S.~B. and {Cunningham}, V. and {Helou}, G. and {Kulkarni}, S.~R. and {Masci}, F.~J. and {Prince}, T. and {Reiley}, D. and {Rodriguez}, H. and {Rusholme}, B. and {Smith}, R.~M. and {Soumagnac}, M.~T.},
        title = "{Candidate Electromagnetic Counterpart to the Binary Black Hole Merger Gravitational-Wave Event S190521g$^{*}$}",
      journal = {\prl},
         year = 2020,
        month = jun,
       volume = {124},
       number = {25},
          eid = {251102},
        pages = {251102},
          doi = {10.1103/PhysRevLett.124.251102},
archivePrefix = {arXiv},
       eprint = {2006.14122},
 primaryClass = {astro-ph.HE},
       adsurl = {https://ui.adsabs.harvard.edu/abs/2020PhRvL.124y1102G}
}

@ARTICLE{BBH_AGN_rates,
       author = {{Ford}, K.~E. Saavik and {McKernan}, Barry},
        title = "{Binary black hole merger rates in AGN discs versus nuclear star clusters: loud beats quiet}",
      journal = {\mnras},
         year = 2022,
        month = dec,
       volume = {517},
       number = {4},
        pages = {5827-5834},
          doi = {10.1093/mnras/stac2861},
archivePrefix = {arXiv},
       eprint = {2109.03212},
 primaryClass = {astro-ph.HE},
       adsurl = {https://ui.adsabs.harvard.edu/abs/2022MNRAS.517.5827F}
}

@ARTICLE{150914Astrophys,
       author = {{Abbott}, B.~P. and {Abbott}, R. and {Abbott}, T.~D. and {Abernathy}, M.~R. and {Acernese}, F. and {Ackley}, K. and {Adams}, C. and {Adams}, T. and {Addesso}, P. and {Adhikari}, R.~X. and {Adya}, V.~B. and {Affeldt}, C. and {Agathos}, M. and {Agatsuma}, K. and {Aggarwal}, N. and {Aguiar}, O.~D. and {Aiello}, L. and {Ain}, A. and {Ajith}, P. and {Allen}, B. and {Allocca}, A. and {Altin}, P.~A. and {Anderson}, S.~B. and {Anderson}, W.~G. and {Arai}, K. and {Araya}, M.~C. and {Arceneaux}, C.~C. and {Areeda}, J.~S. and {Arnaud}, N. and {Arun}, K.~G. and {Ascenzi}, S. and {Ashton}, G. and {Ast}, M. and {Aston}, S.~M. and {Astone}, P. and {Aufmuth}, P. and {Aulbert}, C. and {Babak}, S. and {Bacon}, P. and {Bader}, M.~K.~M. and {Baker}, P.~T. and {Baldaccini}, F. and {Ballardin}, G. and {Ballmer}, S.~W. and {Barayoga}, J.~C. and {Barclay}, S.~E. and {Barish}, B.~C. and {Barker}, D. and {Barone}, F. and {Barr}, B. and {Barsotti}, L. and {Barsuglia}, M. and {Barta}, D. and {Bartlett}, J. and {Bartos}, I. and {Bassiri}, R. and {Basti}, A. and {Batch}, J.~C. and {Baune}, C. and {Bavigadda}, V. and {Bazzan}, M. and {Behnke}, B. and {Bejger}, M. and {Belczynski}, C. and {Bell}, A.~S. and {Bell}, C.~J. and {Berger}, B.~K. and {Bergman}, J. and {Bergmann}, G. and {Berry}, C.~P.~L. and {Bersanetti}, D. and {Bertolini}, A. and {Betzwieser}, J. and {Bhagwat}, S. and {Bhandare}, R. and {Bilenko}, I.~A. and {Billingsley}, G. and {Birch}, J. and {Birney}, R. and {Biscans}, S. and {Bisht}, A. and {Bitossi}, M. and {Biwer}, C. and {Bizouard}, M.~A. and {Blackburn}, J.~K. and {Blair}, C.~D. and {Blair}, D.~G. and {Blair}, R.~M. and {Bloemen}, S. and {Bock}, O. and {Bodiya}, T.~P. and {Boer}, M. and {Bogaert}, G. and {Bogan}, C. and {Bohe}, A. and {Bojtos}, P. and {Bond}, C. and {Bondu}, F. and {Bonnand}, R. and {Boom}, B.~A. and {Bork}, R. and {Boschi}, V. and {Bose}, S. and {Bouffanais}, Y. and {Bozzi}, A. and {Bradaschia}, C. and {Brady}, P.~R. and {Braginsky}, V.~B. and {Branchesi}, M. and {Brau}, J.~E. and {Briant}, T. and {Brillet}, A. and {Brinkmann}, M. and {Brisson}, V. and {Brockill}, P. and {Brooks}, A.~F. and {Brown}, D.~A. and {Brown}, D.~D. and {Brown}, N.~M. and {Buchanan}, C.~C. and {Buikema}, A. and {Bulik}, T. and {Bulten}, H.~J. and {Buonanno}, A. and {Buskulic}, D. and {Buy}, C. and {Byer}, R.~L. and {Cadonati}, L. and {Cagnoli}, G. and {Cahillane}, C. and {Calder{\'o}n Bustillo}, J. and {Callister}, T. and {Calloni}, E. and {Camp}, J.~B. and {Cannon}, K.~C. and {Cao}, J. and {Capano}, C.~D. and {Capocasa}, E. and {Carbognani}, F. and {Caride}, S. and {Casanueva Diaz}, J. and {Casentini}, C. and {Caudill}, S. and {Cavagli{\`a}}, M. and {Cavalier}, F. and {Cavalieri}, R. and {Cella}, G. and {Cepeda}, C. and {Cerboni Baiardi}, L. and {Cerretani}, G. and {Cesarini}, E. and {Chakraborty}, R. and {Chalermsongsak}, T. and {Chamberlin}, S.~J. and {Chan}, M. and {Chao}, S. and {Charlton}, P. and {Chassande-Mottin}, E. and {Chen}, H.~Y. and {Chen}, Y. and {Cheng}, C. and {Chincarini}, A. and {Chiummo}, A. and {Cho}, H.~S. and {Cho}, M. and {Chow}, J.~H. and {Christensen}, N. and {Chu}, Q. and {Chua}, S. and {Chung}, S. and {Ciani}, G. and {Clara}, F. and {Clark}, J.~A. and {Cleva}, F. and {Coccia}, E. and {Cohadon}, P. -F. and {Colla}, A. and {Collette}, C.~G. and {Cominsky}, L. and {Constancio}, Jr., M. and {Conte}, A. and {Conti}, L. and {Cook}, D. and {Corbitt}, T.~R. and {Cornish}, N. and {Corsi}, A. and {Cortese}, S. and {Costa}, C.~A. and {Coughlin}, M.~W. and {Coughlin}, S.~B. and {Coulon}, J. -P. and {Countryman}, S.~T. and {Couvares}, P. and {Cowan}, E.~E. and {Coward}, D.~M. and {Cowart}, M.~J. and {Coyne}, D.~C. and {Coyne}, R. and {Craig}, K. and {Creighton}, J.~D.~E.},
        title = "{Astrophysical Implications of the Binary Black-hole Merger GW150914}",
      journal = {\apjl},
         year = 2016,
        month = feb,
       volume = {818},
       number = {2},
          eid = {L22},
        pages = {L22},
          doi = {10.3847/2041-8205/818/2/L22},
archivePrefix = {arXiv},
       eprint = {1602.03846},
 primaryClass = {astro-ph.HE},
       adsurl = {https://ui.adsabs.harvard.edu/abs/2016ApJ...818L..22A}
}

@ARTICLE{LIGO150914,
       author = {{Abbott}, B.~P. and {Abbott}, R. and {Abbott}, T.~D. and {Abernathy}, M.~R. and {Acernese}, F. and {Ackley}, K. and {Adams}, C. and {Adams}, T. and {Addesso}, P. and {Adhikari}, R.~X. and {Adya}, V.~B. and {Affeldt}, C. and {Agathos}, M. and {Agatsuma}, K. and {Aggarwal}, N. and {Aguiar}, O.~D. and {Aiello}, L. and {Ain}, A. and {Ajith}, P. and {Allen}, B. and {Allocca}, A. and {Altin}, P.~A. and {Anderson}, S.~B. and {Anderson}, W.~G. and {Arai}, K. and {Araya}, M.~C. and {Arceneaux}, C.~C. and {Areeda}, J.~S. and {Arnaud}, N. and {Arun}, K.~G. and {Ascenzi}, S. and {Ashton}, G. and {Ast}, M. and {Aston}, S.~M. and {Astone}, P. and {Aufmuth}, P. and {Aulbert}, C. and {Babak}, S. and {Bacon}, P. and {Bader}, M.~K.~M. and {Baldaccini}, F. and {Ballardin}, G. and {Ballmer}, S.~W. and {Barayoga}, J.~C. and {Barclay}, S.~E. and {Barish}, B.~C. and {Barker}, D. and {Barone}, F. and {Barr}, B. and {Barsotti}, L. and {Barsuglia}, M. and {Barta}, D. and {Bartlett}, J. and {Bartos}, I. and {Bassiri}, R. and {Basti}, A. and {Batch}, J.~C. and {Baune}, C. and {Bavigadda}, V. and {Bazzan}, M. and {Bejger}, M. and {Bell}, A.~S. and {Bergmann}, G. and {Berry}, C.~P.~L. and {Bersanetti}, D. and {Bertolini}, A. and {Betzwieser}, J. and {Bhagwat}, S. and {Bhandare}, R. and {Bilenko}, I.~A. and {Billingsley}, G. and {Birch}, J. and {Birney}, I.~A. and {Birnholtz}, O. and {Biscans}, S. and {Bisht}, A. and {Bitossi}, M. and {Biwer}, C. and {Bizouard}, M.~A. and {Blackburn}, J.~K. and {Blair}, C.~D. and {Blair}, D.~G. and {Blair}, R.~M. and {Bloemen}, S. and {Bock}, O. and {Boer}, M. and {Bogaert}, G. and {Bogan}, C. and {Bohe}, A. and {Bond}, C. and {Bondu}, F. and {Bonnand}, R. and {Boom}, B.~A. and {Bork}, R. and {Boschi}, V. and {Bose}, S. and {Bouffanais}, Y. and {Bozzi}, A. and {Bradaschia}, C. and {Braginsky}, V.~B. and {Branchesi}, M. and {Brau}, J.~E. and {Briant}, T. and {Brillet}, A. and {Brinkmann}, M. and {Brisson}, V. and {Brockill}, P. and {Broida}, J.~E. and {Brooks}, A.~F. and {Brown}, D.~A. and {Brown}, D.~D. and {Brown}, N.~M. and {Brunett}, S. and {Buchanan}, C.~C. and {Buikema}, A. and {Bulik}, T. and {Bulten}, H.~J. and {Buonanno}, A. and {Buskulic}, D. and {Buy}, C. and {Byer}, R.~L. and {Cabero}, M. and {Cadonati}, L. and {Cagnoli}, G. and {Cahillane}, C. and {Bustillo}, J. Calder{\'o}n and {Callister}, T. and {Calloni}, E. and {Camp}, J.~B. and {Cannon}, K.~C. and {Cao}, J. and {Capano}, C.~D. and {Capocasa}, E. and {Carbognani}, F. and {Caride}, S. and {Casanueva Diaz}, J. and {Casentini}, C. and {Caudill}, S. and {Cavagli{\`a}}, M. and {Cavalier}, F. and {Cavalieri}, R. and {Cella}, G. and {Cepeda}, C.~B. and {Cerboni Baiardi}, L. and {Cerretani}, G. and {Cesarini}, E. and {Chamberlin}, S.~J. and {Chan}, M. and {Chao}, S. and {Charlton}, P. and {Chassande-Mottin}, E. and {Chen}, H.~Y. and {Chen}, Y. and {Cheng}, C. and {Chincarini}, A. and {Chiummo}, A. and {Cho}, H.~S. and {Cho}, M. and {Chow}, J.~H. and {Christensen}, N. and {Chu}, Q. and {Chua}, S. and {Chung}, S. and {Ciani}, G. and {Clara}, F. and {Clark}, J.~A. and {Cleva}, F. and {Coccia}, E. and {Cohadon}, P. -F. and {Colla}, A. and {Collette}, C.~G. and {Cominsky}, L. and {Constancio}, Jr., M. and {Conte}, A. and {Conti}, L. and {Cook}, D. and {Corbitt}, T.~R. and {Corsi}, A. and {Cortese}, S. and {Costa}, C.~A. and {Coughlin}, M.~W. and {Coughlin}, S.~B. and {Coulon}, J. -P. and {Countryman}, S.~T. and {Couvares}, P. and {Cowan}, E.~E. and {Coward}, D.~M. and {Cowart}, M.~J. and {Coyne}, D.~C. and {Coyne}, R. and {Craig}, K. and {Creighton}, J.~D.~E. and {Cripe}, J. and {Crowder}, S.~G. and {Cumming}, A. and {Cunningham}, L. and {Cuoco}, E. and {Dal Canton}, T. and {Danilishin}, S.~L. and {D'Antonio}, S.},
        title = "{The basic physics of the binary black hole merger GW150914}",
      journal = {Annalen der Physik},
         year = 2017,
        month = jan,
       volume = {529},
       number = {1-2},
          eid = {1600209},
        pages = {1600209},
          doi = {10.1002/andp.201600209},
archivePrefix = {arXiv},
       eprint = {1608.01940},
 primaryClass = {gr-qc},
       adsurl = {https://ui.adsabs.harvard.edu/abs/2017AnP...52900209A}
}

@ARTICLE{AGNCounterpartModel,
       author = {{Tagawa}, Hiromichi and {Kimura}, Shigeo S. and {Haiman}, Zolt{\'a}n and {Perna}, Rosalba and {Bartos}, Imre},
        title = "{Observable Signature of Merging Stellar-mass Black Holes in Active Galactic Nuclei}",
      journal = {\apj},
         year = 2023,
        month = jun,
       volume = {950},
       number = {1},
          eid = {13},
        pages = {13},
          doi = {10.3847/1538-4357/acc4bb},
archivePrefix = {arXiv},
       eprint = {2301.07111},
 primaryClass = {astro-ph.HE},
       adsurl = {https://ui.adsabs.harvard.edu/abs/2023ApJ...950...13T}
}

@ARTICLE{AGNgasmodel,
       author = {{Bartos}, Imre and {Kocsis}, Bence and {Haiman}, Zolt{\'a}n and {M{\'a}rka}, Szabolcs},
        title = "{Rapid and Bright Stellar-mass Binary Black Hole Mergers in Active Galactic Nuclei}",
      journal = {\apj},
         year = 2017,
        month = feb,
       volume = {835},
       number = {2},
          eid = {165},
        pages = {165},
          doi = {10.3847/1538-4357/835/2/165},
archivePrefix = {arXiv},
       eprint = {1602.03831},
 primaryClass = {astro-ph.HE},
       adsurl = {https://ui.adsabs.harvard.edu/abs/2017ApJ...835..165B}
}

@ARTICLE{S241125nSummary,
       author = {{Zhang}, Shu-Rui and {Wang}, Yu and {Yuan}, Ye-Fei and {Tagawa}, Hiromichi and {Wei}, Yun-Feng and {Li}, Liang and {Cai}, Rong-Gen},
        title = "{S241125n: Binary Black Hole Merger Produces Short GRB in AGN Disk}",
      journal = {arXiv e-prints},
         year = 2025,
        month = may,
          eid = {arXiv:2505.10395},
        pages = {arXiv:2505.10395},
          doi = {10.48550/arXiv.2505.10395},
archivePrefix = {arXiv},
       eprint = {2505.10395},
 primaryClass = {astro-ph.HE},
       adsurl = {https://ui.adsabs.harvard.edu/abs/2025arXiv250510395Z}
}

@article{GUANO_Main,
   title={Gamma-Ray Urgent Archiver for Novel Opportunities (GUANO): Swift/BAT Event Data Dumps on Demand to Enable Sensitive Subthreshold GRB Searches},
   volume={900},
   ISSN={1538-4357},
   url={http://dx.doi.org/10.3847/1538-4357/aba94f},
   DOI={10.3847/1538-4357/aba94f},
   number={1},
   journal={The Astrophysical Journal},
   publisher={American Astronomical Society},
   author={Tohuvavohu, Aaron and Kennea, Jamie A. and DeLaunay, James and Palmer, David M. and Cenko, S. Bradley and Barthelmy, Scott},
   year={2020},
   month=aug, pages={35} }

@article{NITRATES,
   title={Harvesting BAT-GUANO with NITRATES (Non-Imaging Transient Reconstruction and Temporal Search): Detecting and Localizing the Faintest Gamma-Ray Bursts with a Likelihood Framework},
   volume={941},
   ISSN={1538-4357},
   url={http://dx.doi.org/10.3847/1538-4357/ac9d38},
   DOI={10.3847/1538-4357/ac9d38},
   number={2},
   journal={The Astrophysical Journal},
   publisher={American Astronomical Society},
   author={DeLaunay, James and Tohuvavohu, Aaron},
   year={2022},
   month=dec, pages={169} }

@ARTICLE{GW170608_VLA,
       author = {{Artkop}, Kyle and {Smith}, Rachel and {Corsi}, Alessandra and {Giacintucci}, Simona and {Peters}, Wendy M. and {Perna}, Rosalba and {Cenko}, S. Bradley and {Clarke}, Tracy E.},
        title = "{Radio Follow-up of a Candidate {\ensuremath{\gamma}}-Ray Transient in the Sky Localization Area of GW170608}",
      journal = {\apj},
         year = 2019,
        month = oct,
       volume = {884},
       number = {1},
          eid = {16},
        pages = {16},
          doi = {10.3847/1538-4357/ab3e03},
archivePrefix = {arXiv},
       eprint = {1903.10660},
 primaryClass = {astro-ph.HE},
       adsurl = {https://ui.adsabs.harvard.edu/abs/2019ApJ...884...16A}
}

@ARTICLE{LIGO_BBH_EM,
       author = {{Perna}, Rosalba and {Chruslinska}, Martyna and {Corsi}, Alessandra and {Belczynski}, Krzysztof},
        title = "{Binary black hole mergers within the LIGO horizon: statistical properties and prospects for detecting electromagnetic counterparts}",
      journal = {\mnras},
         year = 2018,
        month = jul,
       volume = {477},
       number = {3},
        pages = {4228-4240},
          doi = {10.1093/mnras/sty814},
archivePrefix = {arXiv},
       eprint = {1708.09402},
 primaryClass = {astro-ph.HE},
       adsurl = {https://ui.adsabs.harvard.edu/abs/2018MNRAS.477.4228P}
}

@ARTICLE{BBH_GRBs,
       author = {{Perna}, Rosalba and {Lazzati}, Davide and {Giacomazzo}, Bruno},
        title = "{Short Gamma-Ray Bursts from the Merger of Two Black Holes}",
      journal = {\apjl},
         year = 2016,
        month = apr,
       volume = {821},
       number = {1},
          eid = {L18},
        pages = {L18},
          doi = {10.3847/2041-8205/821/1/L18},
archivePrefix = {arXiv},
       eprint = {1602.05140},
 primaryClass = {astro-ph.HE},
       adsurl = {https://ui.adsabs.harvard.edu/abs/2016ApJ...821L..18P}
}

@ARTICLE{EM_BBH_Limits,
       author = {{Perna}, Rosalba and {Lazzati}, Davide and {Farr}, Will},
        title = "{Limits on Electromagnetic Counterparts of Gravitational-wave-detected Binary Black Hole Mergers}",
      journal = {\apj},
         year = 2019,
        month = apr,
       volume = {875},
       number = {1},
          eid = {49},
        pages = {49},
          doi = {10.3847/1538-4357/ab107b},
archivePrefix = {arXiv},
       eprint = {1901.04522},
 primaryClass = {astro-ph.HE},
       adsurl = {https://ui.adsabs.harvard.edu/abs/2019ApJ...875...49P}
}

@ARTICLE{minidisk_mass,
       author = {{Khan}, Abid and {Paschalidis}, Vasileios and {Ruiz}, Milton and {Shapiro}, Stuart L.},
        title = "{Disks around merging binary black holes: From GW150914 to supermassive black holes}",
      journal = {\prd},
         year = 2018,
        month = feb,
       volume = {97},
       number = {4},
          eid = {044036},
        pages = {044036},
          doi = {10.1103/PhysRevD.97.044036},
archivePrefix = {arXiv},
       eprint = {1801.02624},
 primaryClass = {astro-ph.HE},
       adsurl = {https://ui.adsabs.harvard.edu/abs/2018PhRvD..97d4036K}
}

@ARTICLE{FermiGW,
       author = {{Connaughton}, V. and {Burns}, E. and {Goldstein}, A. and {Blackburn}, L. and {Briggs}, M.~S. and {Zhang}, B. -B. and {Camp}, J. and {Christensen}, N. and {Hui}, C.~M. and {Jenke}, P. and {Littenberg}, T. and {McEnery}, J.~E. and {Racusin}, J. and {Shawhan}, P. and {Singer}, L. and {Veitch}, J. and {Wilson-Hodge}, C.~A. and {Bhat}, P.~N. and {Bissaldi}, E. and {Cleveland}, W. and {Fitzpatrick}, G. and {Giles}, M.~M. and {Gibby}, M.~H. and {von Kienlin}, A. and {Kippen}, R.~M. and {McBreen}, S. and {Mailyan}, B. and {Meegan}, C.~A. and {Paciesas}, W.~S. and {Preece}, R.~D. and {Roberts}, O.~J. and {Sparke}, L. and {Stanbro}, M. and {Toelge}, K. and {Veres}, P.},
        title = "{Fermi GBM Observations of LIGO Gravitational-wave Event GW150914}",
      journal = {\apjl},
         year = 2016,
        month = jul,
       volume = {826},
       number = {1},
          eid = {L6},
        pages = {L6},
          doi = {10.3847/2041-8205/826/1/L6},
archivePrefix = {arXiv},
       eprint = {1602.03920},
 primaryClass = {astro-ph.HE},
       adsurl = {https://ui.adsabs.harvard.edu/abs/2016ApJ...826L...6C}
}

@article{BBHLigo1,
  title = {GW151226: Observation of Gravitational Waves from a 22-Solar-Mass Binary Black Hole Coalescence},
  author = {Abbott, B. P. and Abbott, R. and Abbott, T. D. and Abernathy, M. R. and Acernese, F. and Ackley, K. and Adams, C. and Adams, T. and Addesso, P. and Adhikari, R. X. and Adya, V. B. and Affeldt, C. and Agathos, M. and Agatsuma, K. and Aggarwal, N. and Aguiar, O. D. and Aiello, L. and Ain, A. and Ajith, P. and Allen, B. and Allocca, A. and Altin, P. A. and Anderson, S. B. and Anderson, W. G. and Arai, K. and Araya, M. C. and Arceneaux, C. C. and Areeda, J. S. and Arnaud, N. and Arun, K. G. and Ascenzi, S. and Ashton, G. and Ast, M. and Aston, S. M. and Astone, P. and Aufmuth, P. and Aulbert, C. and Babak, S. and Bacon, P. and Bader, M. K. M. and Baker, P. T. and Baldaccini, F. and Ballardin, G. and Ballmer, S. W. and Barayoga, J. C. and Barclay, S. E. and Barish, B. C. and Barker, D. and Barone, F. and Barr, B. and Barsotti, L. and Barsuglia, M. and Barta, D. and Bartlett, J. and Bartos, I. and Bassiri, R. and Basti, A. and Batch, J. C. and Baune, C. and Bavigadda, V. and Bazzan, M. and Bejger, M. and Bell, A. S. and Berger, B. K. and Bergmann, G. and Berry, C. P. L. and Bersanetti, D. and Bertolini, A. and Betzwieser, J. and Bhagwat, S. and Bhandare, R. and Bilenko, I. A. and Billingsley, G. and Birch, J. and Birney, I. A. and Birnholtz, O. and Biscans, S. and Bisht, A. and Bitossi, M. and Biwer, C. and Bizouard, M. A. and Blackburn, J. K. and Blair, C. D. and Blair, D. G. and Blair, R. M. and Bloemen, S. and Bock, O. and Boer, M. and Bogaert, G. and Bogan, C. and Bohe, A. and Bond, C. and Bondu, F. and Bonnand, R. and Boom, B. A. and Bork, R. and Boschi, V. and Bose, S. and Bouffanais, Y. and Bozzi, A. and Bradaschia, C. and Brady, P. R. and Braginsky, V. B. and Branchesi, M. and Brau, J. E. and Briant, T. and Brillet, A. and Brinkmann, M. and Brisson, V. and Brockill, P. and Broida, J. E. and Brooks, A. F. and Brown, D. A. and Brown, D. D. and Brown, N. M. and Brunett, S. and Buchanan, C. C. and Buikema, A. and Bulik, T. and Bulten, H. J. and Buonanno, A. and Buskulic, D. and Buy, C. and Byer, R. L. and Cabero, M. and Cadonati, L. and Cagnoli, G. and Cahillane, C. and Calder\'on Bustillo, J. and Callister, T. and Calloni, E. and Camp, J. B. and Cannon, K. C. and Cao, J. and Capano, C. D. and Capocasa, E. and Carbognani, F. and Caride, S. and Casanueva Diaz, J. and Casentini, C. and Caudill, S. and Cavagli\`a, M. and Cavalier, F. and Cavalieri, R. and Cella, G. and Cepeda, C. B. and Cerboni Baiardi, L. and Cerretani, G. and Cesarini, E. and Chamberlin, S. J. and Chan, M. and Chao, S. and Charlton, P. and Chassande-Mottin, E. and Cheeseboro, B. D. and Chen, H. Y. and Chen, Y. and Cheng, C. and Chincarini, A. and Chiummo, A. and Cho, H. S. and Cho, M. and Chow, J. H. and Christensen, N. and Chu, Q. and Chua, S. and Chung, S. and Ciani, G. and Clara, F. and Clark, J. A. and Cleva, F. and Coccia, E. and Cohadon, P.-F. and Colla, A. and Collette, C. G. and Cominsky, L. and Constancio, M. and Conte, A. and Conti, L. and Cook, D. and Corbitt, T. R. and Cornish, N. and Corsi, A. and Cortese, S. and Costa, C. A. and Coughlin, M. W. and Coughlin, S. B. and Coulon, J.-P. and Countryman, S. T. and Couvares, P. and Cowan, E. E. and Coward, D. M. and Cowart, M. J. and Coyne, D. C. and Coyne, R. and Craig, K. and Creighton, J. D. E. and Cripe, J. and Crowder, S. G. and Cumming, A. and Cunningham, L. and Cuoco, E. and Dal Canton, T. and Danilishin, S. L. and D'Antonio, S. and Danzmann, K. and Darman, N. S. and Dasgupta, A. and Da Silva Costa, C. F. and Dattilo, V. and Dave, I. and Davier, M. and Davies, G. S. and Daw, E. J. and Day, R. and De, S. and DeBra, D. and Debreczeni, G. and Degallaix, J. and De Laurentis, M. and Del\'eglise, S. and Del Pozzo, W. and Denker, T. and Dent, T. and Dergachev, V. and De Rosa, R. and DeRosa, R. T. and DeSalvo, R. and Devine, R. C. and Dhurandhar, S. and D\'{\i}az, M. C. and Di Fiore, L. and Di Giovanni, M. and Di Girolamo, T. and Di Lieto, A. and Di Pace, S. and Di Palma, I. and Di Virgilio, A. and Dolique, V. and Donovan, F. and Dooley, K. L. and Doravari, S. and Douglas, R. and Downes, T. P. and Drago, M. and Drever, R. W. P. and Driggers, J. C. and Ducrot, M. and Dwyer, S. E. and Edo, T. B. and Edwards, M. C. and Effler, A. and Eggenstein, H.-B. and Ehrens, P. and Eichholz, J. and Eikenberry, S. S. and Engels, W. and Essick, R. C. and Etzel, T. and Evans, M. and Evans, T. M. and Everett, R. and Factourovich, M. and Fafone, V. and Fair, H. and Fairhurst, S. and Fan, X. and Fang, Q. and Farinon, S. and Farr, B. and Farr, W. M. and Favata, M. and Fays, M. and Fehrmann, H. and Fejer, M. M. and Fenyvesi, E. and Ferrante, I. and Ferreira, E. C. and Ferrini, F. and Fidecaro, F. and Fiori, I. and Fiorucci, D. and Fisher, R. P. and Flaminio, R. and Fletcher, M. and Fong, H. and Fournier, J.-D. and Frasca, S. and Frasconi, F. and Frei, Z. and Freise, A. and Frey, R. and Frey, V. and Fritschel, P. and Frolov, V. V. and Fulda, P. and Fyffe, M. and Gabbard, H. A. G. and Gair, J. R. and Gammaitoni, L. and Gaonkar, S. G. and Garufi, F. and Gaur, G. and Gehrels, N. and Gemme, G. and Geng, P. and Genin, E. and Gennai, A. and George, J. and Gergely, L. and Germain, V. and Ghosh, Abhirup and Ghosh, Archisman and Ghosh, S. and Giaime, J. A. and Giardina, K. D. and Giazotto, A. and Gill, K. and Glaefke, A. and Goetz, E. and Goetz, R. and Gondan, L. and Gonz\'alez, G. and Gonzalez Castro, J. M. and Gopakumar, A. and Gordon, N. A. and Gorodetsky, M. L. and Gossan, S. E. and Gosselin, M. and Gouaty, R. and Grado, A. and Graef, C. and Graff, P. B. and Granata, M. and Grant, A. and Gras, S. and Gray, C. and Greco, G. and Green, A. C. and Groot, P. and Grote, H. and Grunewald, S. and Guidi, G. M. and Guo, X. and Gupta, A. and Gupta, M. K. and Gushwa, K. E. and Gustafson, E. K. and Gustafson, R. and Hacker, J. J. and Hall, B. R. and Hall, E. D. and Hamilton, H. and Hammond, G. and Haney, M. and Hanke, M. M. and Hanks, J. and Hanna, C. and Hannam, M. D. and Hanson, J. and Hardwick, T. and Harms, J. and Harry, G. M. and Harry, I. W. and Hart, M. J. and Hartman, M. T. and Haster, C.-J. and Haughian, K. and Healy, J. and Heidmann, A. and Heintze, M. C. and Heitmann, H. and Hello, P. and Hemming, G. and Hendry, M. and Heng, I. S. and Hennig, J. and Henry, J. and Heptonstall, A. W. and Heurs, M. and Hild, S. and Hoak, D. and Hofman, D. and Holt, K. and Holz, D. E. and Hopkins, P. and Hough, J. and Houston, E. A. and Howell, E. J. and Hu, Y. M. and Huang, S. and Huerta, E. A. and Huet, D. and Hughey, B. and Husa, S. and Huttner, S. H. and Huynh-Dinh, T. and Indik, N. and Ingram, D. R. and Inta, R. and Isa, H. N. and Isac, J.-M. and Isi, M. and Isogai, T. and Iyer, B. R. and Izumi, K. and Jacqmin, T. and Jang, H. and Jani, K. and Jaranowski, P. and Jawahar, S. and Jian, L. and Jim\'enez-Forteza, F. and Johnson, W. W. and Johnson-McDaniel, N. K. and Jones, D. I. and Jones, R. and Jonker, R. J. G. and Ju, L. and K, Haris and Kalaghatgi, C. V. and Kalogera, V. and Kandhasamy, S. and Kang, G. and Kanner, J. B. and Kapadia, S. J. and Karki, S. and Karvinen, K. S. and Kasprzack, M. and Katsavounidis, E. and Katzman, W. and Kaufer, S. and Kaur, T. and Kawabe, K. and K\'ef\'elian, F. and Kehl, M. S. and Keitel, D. and Kelley, D. B. and Kells, W. and Kennedy, R. and Key, J. S. and Khalili, F. Y. and Khan, I. and Khan, S. and Khan, Z. and Khazanov, E. A. and Kijbunchoo, N. and Kim, Chi-Woong and Kim, Chunglee and Kim, J. and Kim, K. and Kim, N. and Kim, W. and Kim, Y.-M. and Kimbrell, S. J. and King, E. J. and King, P. J. and Kissel, J. S. and Klein, B. and Kleybolte, L. and Klimenko, S. and Koehlenbeck, S. M. and Koley, S. and Kondrashov, V. and Kontos, A. and Korobko, M. and Korth, W. Z. and Kowalska, I. and Kozak, D. B. and Kringel, V. and Krishnan, B. and Kr\'olak, A. and Krueger, C. and Kuehn, G. and Kumar, P. and Kumar, R. and Kuo, L. and Kutynia, A. and Lackey, B. D. and Landry, M. and Lange, J. and Lantz, B. and Lasky, P. D. and Laxen, M. and Lazzarini, A. and Lazzaro, C. and Leaci, P. and Leavey, S. and Lebigot, E. O. and Lee, C. H. and Lee, H. K. and Lee, H. M. and Lee, K. and Lenon, A. and Leonardi, M. and Leong, J. R. and Leroy, N. and Letendre, N. and Levin, Y. and Lewis, J. B. and Li, T. G. F. and Libson, A. and Littenberg, T. B. and Lockerbie, N. A. and Lombardi, A. L. and London, L. T. and Lord, J. E. and Lorenzini, M. and Loriette, V. and Lormand, M. and Losurdo, G. and Lough, J. D. and Lousto, C. O. and L\"uck, H. and Lundgren, A. P. and Lynch, R. and Ma, Y. and Machenschalk, B. and MacInnis, M. and Macleod, D. M. and Maga\~na-Sandoval, F. and Maga\~na Zertuche, L. and Magee, R. M. and Majorana, E. and Maksimovic, I. and Malvezzi, V. and Man, N. and Mandel, I. and Mandic, V. and Mangano, V. and Mansell, G. L. and Manske, M. and Mantovani, M. and Marchesoni, F. and Marion, F. and M\'arka, S. and M\'arka, Z. and Markosyan, A. S. and Maros, E. and Martelli, F. and Martellini, L. and Martin, I. W. and Martynov, D. V. and Marx, J. N. and Mason, K. and Masserot, A. and Massinger, T. J. and Masso-Reid, M. and Mastrogiovanni, S. and Matichard, F. and Matone, L. and Mavalvala, N. and Mazumder, N. and McCarthy, R. and McClelland, D. E. and McCormick, S. and McGuire, S. C. and McIntyre, G. and McIver, J. and McManus, D. J. and McRae, T. and McWilliams, S. T. and Meacher, D. and Meadors, G. D. and Meidam, J. and Melatos, A. and Mendell, G. and Mercer, R. A. and Merilh, E. L. and Merzougui, M. and Meshkov, S. and Messenger, C. and Messick, C. and Metzdorff, R. and Meyers, P. M. and Mezzani, F. and Miao, H. and Michel, C. and Middleton, H. and Mikhailov, E. E. and Milano, L. and Miller, A. L. and Miller, A. and Miller, B. B. and Miller, J. and Millhouse, M. and Minenkov, Y. and Ming, J. and Mirshekari, S. and Mishra, C. and Mitra, S. and Mitrofanov, V. P. and Mitselmakher, G. and Mittleman, R. and Moggi, A. and Mohan, M. and Mohapatra, S. R. P. and Montani, M. and Moore, B. C. and Moore, C. J. and Moraru, D. and Moreno, G. and Morriss, S. R. and Mossavi, K. and Mours, B. and Mow-Lowry, C. M. and Mueller, G. and Muir, A. W. and Mukherjee, Arunava and Mukherjee, D. and Mukherjee, S. and Mukund, N. and Mullavey, A. and Munch, J. and Murphy, D. J. and Murray, P. G. and Mytidis, A. and Nardecchia, I. and Naticchioni, L. and Nayak, R. K. and Nedkova, K. and Nelemans, G. and Nelson, T. J. N. and Neri, M. and Neunzert, A. and Newton, G. and Nguyen, T. T. and Nielsen, A. B. and Nissanke, S. and Nitz, A. and Nocera, F. and Nolting, D. and Normandin, M. E. N. and Nuttall, L. K. and Oberling, J. and Ochsner, E. and O'Dell, J. and Oelker, E. and Ogin, G. H. and Oh, J. J. and Oh, S. H. and Ohme, F. and Oliver, M. and Oppermann, P. and Oram, Richard J. and O'Reilly, B. and O'Shaughnessy, R. and Ottaway, D. J. and Overmier, H. and Owen, B. J. and Pai, A. and Pai, S. A. and Palamos, J. R. and Palashov, O. and Palomba, C. and Pal-Singh, A. and Pan, H. and Pankow, C. and Pannarale, F. and Pant, B. C. and Paoletti, F. and Paoli, A. and Papa, M. A. and Paris, H. R. and Parker, W. and Pascucci, D. and Pasqualetti, A. and Passaquieti, R. and Passuello, D. and Patricelli, B. and Patrick, Z. and Pearlstone, B. L. and Pedraza, M. and Pedurand, R. and Pekowsky, L. and Pele, A. and Penn, S. and Perreca, A. and Perri, L. M. and Pfeiffer, H. P. and Phelps, M. and Piccinni, O. J. and Pichot, M. and Piergiovanni, F. and Pierro, V. and Pillant, G. and Pinard, L. and Pinto, I. M. and Pitkin, M. and Poe, M. and Poggiani, R. and Popolizio, P. and Post, A. and Powell, J. and Prasad, J. and Predoi, V. and Prestegard, T. and Price, L. R. and Prijatelj, M. and Principe, M. and Privitera, S. and Prix, R. and Prodi, G. A. and Prokhorov, L. and Puncken, O. and Punturo, M. and Puppo, P. and P\"urrer, M. and Qi, H. and Qin, J. and Qiu, S. and Quetschke, V. and Quintero, E. A. and Quitzow-James, R. and Raab, F. J. and Rabeling, D. S. and Radkins, H. and Raffai, P. and Raja, S. and Rajan, C. and Rakhmanov, M. and Rapagnani, P. and Raymond, V. and Razzano, M. and Re, V. and Read, J. and Reed, C. M. and Regimbau, T. and Rei, L. and Reid, S. and Reitze, D. H. and Rew, H. and Reyes, S. D. and Ricci, F. and Riles, K. and Rizzo, M. and Robertson, N. A. and Robie, R. and Robinet, F. and Rocchi, A. and Rolland, L. and Rollins, J. G. and Roma, V. J. and Romano, J. D. and Romano, R. and Romanov, G. and Romie, J. H. and Rosi\ifmmode \acute{n}\else \'{n}\fi{}ska, D. and Rowan, S. and R\"udiger, A. and Ruggi, P. and Ryan, K. and Sachdev, S. and Sadecki, T. and Sadeghian, L. and Sakellariadou, M. and Salconi, L. and Saleem, M. and Salemi, F. and Samajdar, A. and Sammut, L. and Sanchez, E. J. and Sandberg, V. and Sandeen, B. and Sanders, J. R. and Sassolas, B. and Sathyaprakash, B. S. and Saulson, P. R. and Sauter, O. E. S. and Savage, R. L. and Sawadsky, A. and Schale, P. and Schilling, R. and Schmidt, J. and Schmidt, P. and Schnabel, R. and Schofield, R. M. S. and Sch\"onbeck, A. and Schreiber, E. and Schuette, D. and Schutz, B. F. and Scott, J. and Scott, S. M. and Sellers, D. and Sengupta, A. S. and Sentenac, D. and Sequino, V. and Sergeev, A. and Setyawati, Y. and Shaddock, D. A. and Shaffer, T. and Shahriar, M. S. and Shaltev, M. and Shapiro, B. and Shawhan, P. and Sheperd, A. and Shoemaker, D. H. and Shoemaker, D. M. and Siellez, K. and Siemens, X. and Sieniawska, M. and Sigg, D. and Silva, A. D. and Singer, A. and Singer, L. P. and Singh, A. and Singh, R. and Singhal, A. and Sintes, A. M. and Slagmolen, B. J. J. and Smith, J. R. and Smith, N. D. and Smith, R. J. E. and Son, E. J. and Sorazu, B. and Sorrentino, F. and Souradeep, T. and Srivastava, A. K. and Staley, A. and Steinke, M. and Steinlechner, J. and Steinlechner, S. and Steinmeyer, D. and Stephens, B. C. and Stevenson, S. P. and Stone, R. and Strain, K. A. and Straniero, N. and Stratta, G. and Strauss, N. A. and Strigin, S. and Sturani, R. and Stuver, A. L. and Summerscales, T. Z. and Sun, L. and Sunil, S. and Sutton, P. J. and Swinkels, B. L. and Szczepa\ifmmode \acute{n}\else \'{n}\fi{}czyk, M. J. and Tacca, M. and Talukder, D. and Tanner, D. B. and T\'apai, M. and Tarabrin, S. P. and Taracchini, A. and Taylor, R. and Theeg, T. and Thirugnanasambandam, M. P. and Thomas, E. G. and Thomas, M. and Thomas, P. and Thorne, K. A. and Thrane, E. and Tiwari, S. and Tiwari, V. and Tokmakov, K. V. and Toland, K. and Tomlinson, C. and Tonelli, M. and Tornasi, Z. and Torres, C. V. and Torrie, C. I. and T\"oyr\"a, D. and Travasso, F. and Traylor, G. and Trifir\`o, D. and Tringali, M. C. and Trozzo, L. and Tse, M. and Turconi, M. and Tuyenbayev, D. and Ugolini, D. and Unnikrishnan, C. S. and Urban, A. L. and Usman, S. A. and Vahlbruch, H. and Vajente, G. and Valdes, G. and Vallisneri, M. and van Bakel, N. and van Beuzekom, M. and van den Brand, J. F. J. and Van Den Broeck, C. and Vander-Hyde, D. C. and van der Schaaf, L. and van Heijningen, J. V. and van Veggel, A. A. and Vardaro, M. and Vass, S. and Vas\'uth, M. and Vaulin, R. and Vecchio, A. and Vedovato, G. and Veitch, J. and Veitch, P. J. and Venkateswara, K. and Verkindt, D. and Vetrano, F. and Vicer\'e, A. and Vinciguerra, S. and Vine, D. J. and Vinet, J.-Y. and Vitale, S. and Vo, T. and Vocca, H. and Vorvick, C. and Voss, D. V. and Vousden, W. D. and Vyatchanin, S. P. and Wade, A. R. and Wade, L. E. and Wade, M. and Walker, M. and Wallace, L. and Walsh, S. and Wang, G. and Wang, H. and Wang, M. and Wang, X. and Wang, Y. and Ward, R. L. and Warner, J. and Was, M. and Weaver, B. and Wei, L.-W. and Weinert, M. and Weinstein, A. J. and Weiss, R. and Wen, L. and We\ss{}els, P. and Westphal, T. and Wette, K. and Whelan, J. T. and Whiting, B. F. and Williams, R. D. and Williamson, A. R. and Willis, J. L. and Willke, B. and Wimmer, M. H. and Winkler, W. and Wipf, C. C. and Wittel, H. and Woan, G. and Woehler, J. and Worden, J. and Wright, J. L. and Wu, D. S. and Wu, G. and Yablon, J. and Yam, W. and Yamamoto, H. and Yancey, C. C. and Yu, H. and Yvert, M. and Zadro\ifmmode \dot{z}\else \.{z}\fi{}ny, A. and Zangrando, L. and Zanolin, M. and Zendri, J.-P. and Zevin, M. and Zhang, L. and Zhang, M. and Zhang, Y. and Zhao, C. and Zhou, M. and Zhou, Z. and Zhu, X. J. and Zucker, M. E. and Zuraw, S. E. and Zweizig, J. and Boyle, M. and Hemberger, D. and Kidder, L. E. and Lovelace, G. and Ossokine, S. and Scheel, M. and Szilagyi, B. and Teukolsky, S.},
  collaboration = {LIGO Scientific Collaboration and Virgo Collaboration},
  journal = {Phys. Rev. Lett.},
  volume = {116},
  issue = {24},
  pages = {241103},
  numpages = {14},
  year = {2016},
  month = {Jun},
  publisher = {American Physical Society},
  doi = {10.1103/PhysRevLett.116.241103},
  url = {https://link.aps.org/doi/10.1103/PhysRevLett.116.241103}
}

@ARTICLE{170817GW,
       author = {{Abbott}, B.~P. and {Abbott}, R. and {Abbott}, T.~D. and {Acernese}, F. and {Ackley}, K. and {Adams}, C. and {Adams}, T. and {Addesso}, P. and {Adhikari}, R.~X. and {Adya}, V.~B. and {Affeldt}, C. and {Afrough}, M. and {Agarwal}, B. and {Agathos}, M. and {Agatsuma}, K. and {Aggarwal}, N. and {Aguiar}, O.~D. and {Aiello}, L. and {Ain}, A. and {Ajith}, P. and {Allen}, B. and {Allen}, G. and {Allocca}, A. and {Altin}, P.~A. and {Amato}, A. and {Ananyeva}, A. and {Anderson}, S.~B. and {Anderson}, W.~G. and {Angelova}, S.~V. and {Antier}, S. and {Appert}, S. and {Arai}, K. and {Araya}, M.~C. and {Areeda}, J.~S. and {Arnaud}, N. and {Arun}, K.~G. and {Ascenzi}, S. and {Ashton}, G. and {Ast}, M. and {Aston}, S.~M. and {Astone}, P. and {Atallah}, D.~V. and {Aufmuth}, P. and {Aulbert}, C. and {AultONeal}, K. and {Austin}, C. and {Avila-Alvarez}, A. and {Babak}, S. and {Bacon}, P. and {Bader}, M.~K.~M. and {Bae}, S. and {Baker}, P.~T. and {Baldaccini}, F. and {Ballardin}, G. and {Ballmer}, S.~W. and {Banagiri}, S. and {Barayoga}, J.~C. and {Barclay}, S.~E. and {Barish}, B.~C. and {Barker}, D. and {Barkett}, K. and {Barone}, F. and {Barr}, B. and {Barsotti}, L. and {Barsuglia}, M. and {Barta}, D. and {Barthelmy}, S.~D. and {Bartlett}, J. and {Bartos}, I. and {Bassiri}, R. and {Basti}, A. and {Batch}, J.~C. and {Bawaj}, M. and {Bayley}, J.~C. and {Bazzan}, M. and {B{\'e}csy}, B. and {Beer}, C. and {Bejger}, M. and {Belahcene}, I. and {Bell}, A.~S. and {Berger}, B.~K. and {Bergmann}, G. and {Bero}, J.~J. and {Berry}, C.~P.~L. and {Bersanetti}, D. and {Bertolini}, A. and {Betzwieser}, J. and {Bhagwat}, S. and {Bhandare}, R. and {Bilenko}, I.~A. and {Billingsley}, G. and {Billman}, C.~R. and {Birch}, J. and {Birney}, R. and {Birnholtz}, O. and {Biscans}, S. and {Biscoveanu}, S. and {Bisht}, A. and {Bitossi}, M. and {Biwer}, C. and {Bizouard}, M.~A. and {Blackburn}, J.~K. and {Blackman}, J. and {Blair}, C.~D. and {Blair}, D.~G. and {Blair}, R.~M. and {Bloemen}, S. and {Bock}, O. and {Bode}, N. and {Boer}, M. and {Bogaert}, G. and {Bohe}, A. and {Bondu}, F. and {Bonilla}, E. and {Bonnand}, R. and {Boom}, B.~A. and {Bork}, R. and {Boschi}, V. and {Bose}, S. and {Bossie}, K. and {Bouffanais}, Y. and {Bozzi}, A. and {Bradaschia}, C. and {Brady}, P.~R. and {Branchesi}, M. and {Brau}, J.~E. and {Briant}, T. and {Brillet}, A. and {Brinkmann}, M. and {Brisson}, V. and {Brockill}, P. and {Broida}, J.~E. and {Brooks}, A.~F. and {Brown}, D.~A. and {Brown}, D.~D. and {Brunett}, S. and {Buchanan}, C.~C. and {Buikema}, A. and {Bulik}, T. and {Bulten}, H.~J. and {Buonanno}, A. and {Buskulic}, D. and {Buy}, C. and {Byer}, R.~L. and {Cabero}, M. and {Cadonati}, L. and {Cagnoli}, G. and {Cahillane}, C. and {Calder{\'o}n Bustillo}, J. and {Callister}, T.~A. and {Calloni}, E. and {Camp}, J.~B. and {Canepa}, M. and {Canizares}, P. and {Cannon}, K.~C. and {Cao}, H. and {Cao}, J. and {Capano}, C.~D. and {Capocasa}, E. and {Carbognani}, F. and {Caride}, S. and {Carney}, M.~F. and {Casanueva Diaz}, J. and {Casentini}, C. and {Caudill}, S. and {Cavagli{\`a}}, M. and {Cavalier}, F. and {Cavalieri}, R. and {Cella}, G. and {Cepeda}, C.~B. and {Cerd{\'a}-Dur{\'a}n}, P. and {Cerretani}, G. and {Cesarini}, E. and {Chamberlin}, S.~J. and {Chan}, M. and {Chao}, S. and {Charlton}, P. and {Chase}, E. and {Chassande-Mottin}, E. and {Chatterjee}, D. and {Chatziioannou}, K. and {Cheeseboro}, B.~D. and {Chen}, H.~Y. and {Chen}, X. and {Chen}, Y. and {Cheng}, H. -P. and {Chia}, H. and {Chincarini}, A. and {Chiummo}, A. and {Chmiel}, T. and {Cho}, H.~S. and {Cho}, M. and {Chow}, J.~H. and {Christensen}, N. and {Chu}, Q. and {Chua}, A.~J.~K. and {Chua}, S. and {Chung}, A.~K.~W. and {Chung}, S. and {Ciani}, G.},
        title = "{Multi-messenger Observations of a Binary Neutron Star Merger}",
      journal = {\apjl},
         year = 2017,
        month = oct,
       volume = {848},
       number = {2},
          eid = {L12},
        pages = {L12},
          doi = {10.3847/2041-8213/aa91c9},
archivePrefix = {arXiv},
       eprint = {1710.05833},
 primaryClass = {astro-ph.HE},
       adsurl = {https://ui.adsabs.harvard.edu/abs/2017ApJ...848L..12A}
}

@misc{GCN38305,
  author       = "{{LIGO Scientific Collaboration} and {Virgo Collaboration}, and {KAGRA Collaboration}}",
  title        = "",
  howpublished = "GCN Circular ",
  month        = "",
  year         = "2024",
  note         = ""
}
\bibliographystyle{aasjournal}

\end{document}